\documentclass[letters,usenatbib]{mnras}

\usepackage{amssymb}	

\usepackage{newtxtext,newtxmath}
\usepackage{soul}

\usepackage[T1]{fontenc}

\DeclareRobustCommand{\VAN}[3]{#2}
\let\VANthebibliography\thebibliography
\def\thebibliography{\DeclareRobustCommand{\VAN}[3]{##3}\VANthebibliography}

\usepackage{graphicx}	
\usepackage{amsmath}	
\usepackage{multirow}

\title[Gyrochronology relations against coeval pairs]{The breakdown of current gyrochronology as evidenced by old coeval stars}

\author[J. Silva-Beyer et al.]{
Joaquín Silva-Beyer,$^{1}$\thanks{E-mail: jasilva11@uc.cl}
Diego Godoy-Rivera,$^{2,3,4}$
Julio Chanamé$^{1}$
\\
$^{1}$Instituto de Astrofísica, Pontificia Universidad Católica de Chile, Av. Vicuña Mackenna 4860, 782-0436 Macul, Santiago, Chile\\
$^{2}$Instituto de Astrofísica de Canarias (IAC), E-38205 La Laguna, Tenerife, Spain \\
$^{3}$Universidad de La Laguna (ULL), Departamento de Astrofísica, E-38206 La Laguna, Tenerife, Spain \\
$^{4}$Department of Astronomy, The Ohio State University, 140 West 18th Avenue, Columbus, OH 43210, USA\\
}
\date{Accepted XXX. Received YYY; in original form ZZZ}

\pubyear{2023}

\begin{document}
\label{firstpage}
\pagerange{\pageref{firstpage}--\pageref{lastpage}}
\maketitle

\begin{abstract}
Gyrochronology can yield useful ages for field main-sequence stars, a regime where other
techniques are problematic. Typically, gyrochronology relations are calibrated using
young ($\lesssim$ 2 Gyr) clusters, but the constraints at older ages are scarce, making them
potentially inaccurate and imprecise. In order to test the performance of existing
relations, we construct samples of stellar pairs with coeval components, for a range of
ages and with available rotation periods. These include randomly paired stars in
clusters, and wide binaries in the Kepler field. We design indicators that, based on the
measured rotation periods and expectations from gyrochronology, quantify the (dis)agreement
between the coeval pairs and the gyrochronology calibrations under scrutiny. Our results
show that wide binaries and cluster members are in better concordance with gyrochronology
than samples of randomly paired field stars, confirming that the relations have predicting
power. However, the agreement with the examined relations decreases for older stars,
revealing a degradation of the examined relations with age, in agreement with recent works.
This highlights the need for novel empirical constraints at older ages that may allow
revised calibrations. Notably, using coeval stars to test gyrochronology poses the advantage of circumventing the need for age determinations while simultaneously exploiting larger samples at older ages. Our test is independent of any specific age-rotation relation, and it can be used to evaluate future spin-down models. In addition, taking gyrochronology at face value, we note that our results provide new empirical evidence that the components of field wide binaries are indeed coeval.

\end{abstract}




\section{Introduction} \label{sec:intro}

One of the most important parameters in stellar astrophysics is the age of a star, as it is essential to the study and understanding of its physical properties, evolution, and structure \citep{2010ARA&A..48..581S}. Furthermore, the determination of stellar ages is crucial to the study of young and old stellar populations in the Galactic field, stellar streams, and moving groups \citep[e.g.,][]{2019AJ....158...77C, 2022AJ....163..275A, 2022A&A...657L...3M, 2023MNRAS.522.1288B, 2022Natur.603..599X, 2022MNRAS.510.4669S}, as well as for the characterization of exoplanet hosts \citep[e.g.,][]{2016A&A...585A...5B, 2020AJ....160..108B,2023MNRAS.520.5283G}. Ages are, however, difficult to determine with precision. Most methods for determining ages rely on coeval sets of stars like clusters \citep[e.g.,][]{2019A&A...623A.108B, 2021ApJS..257...46G}, stars suitable for asteroseismic analysis \citep[e.g.,][]{2014ApJ...790L..23D, 2014ApJS..210....1C, 2018ApJS..239...32P}, or evolved stars for which isochrone fitting is sufficiently precise, like turnoff stars or subgiants (e.g., \citealt{2015A&A...577A..90M}; \citealt{2017ApJS..232....2X}; \citealt{2021ApJ...923..181B}; \citealt{ 2021ApJ...915...19G}). On the other hand, the more numerous low-mass main sequence (MS) stars in the field, pose a difficult case. By being in the field, they are not obviously associated to a coeval set of stars that could help anchor their age, as in a cluster.  Moreover, since their location on the Hertzsprung-Russell diagram (HRD) barely changes over Gyr timescales, isochrone fitting is usually imprecise in most cases.  Additionally, due to both technical and astrophysical reasons, the number of MS stars with asteroseismic ages is still limited \citep[][]{2022A&A...657A..31M}. Galactic kinematics might be an option for this kind of stars, although likely restricted to populations rather than individual targets \citep[e.g.,][]{2016ApJ...821...93N,2020AJ....160...90A}.

One technique that aims to solve this problem is gyrochronology \citep{2003ApJ...586..464B}. It relies on the fact that stars with convective envelopes experience angular momentum losses due to magnetized winds, which results in a spin-down of their rotation rates \citep[][]{1958ApJ...128..664P, 1962AnAp...25...18S, 1988ApJ...333..236K}. The influential work of \citet{1972ApJ...171..565S} first showed that rotational velocity decays as $\propto \text{age}^{-1/2}$ for G-type stars. More recent works have calibrated gyrochronology with F, G, K, and early M stars \citep[e.g.,][]{2007ApJ...669.1167B,  2008ApJ...687.1264M}, showing that age-rotation relations depend critically on mass (or color as a proxy for mass). During the last decade, empirical relations have been developed with increasing precision \citep[e.g.,][]{2014A&A...572A..34G, 2015MNRAS.450.1787A,2019AJ....158..173A,2023ApJ...947L...3B}.

Most empirical studies of gyrochronology have used star clusters as calibrators due to the coeval nature of their members \citep[e.g.,][]{2009ApJ...695..679M, 2015Natur.517..589M,2019ApJ...879...49C}. However, this poses a few problems. First, the calibration can only be done in clusters with sufficiently large samples of stars with measured rotation periods, which are not very common. Second, most of these clusters are relatively young ($\lesssim 2$ Gyr), making the application of gyrochronology to older stars uncertain, as the relations are not well calibrated in this regime and may deviate from the classical Skumanich-like dependence \citep[][]{2019ApJ...871...39M}. To mitigate these issues, one would ideally also calibrate gyrochronology with field MS stars, but this is difficult due to the previously discussed issues in determining their ages. In an attempt to solve some of these problems, a different approach can be taken: testing and calibrating gyrochronology with wide binaries.

Unlike clusters, wide binaries in the solar neighborhood span basically the full age and metallicity distributions of stars in the Galaxy.  Quite crucially, different formation channels for wide binaries all imply that the components of any given pair must be stars of common origins, so that they are expected to be coeval \citep{2010MNRAS.404.1835K,2010ApJ...725.1485O,2011MNRAS.415.1179M,2016MNRAS.459.4499E,2017NatAs...1E.172L,2021MNRAS.501.3670P, 2022MNRAS.512.3383H}.  This is supported by empirical studies of binaries in young star forming regions (\citealt{2009ApJ...704..531K}), and older field stars (\citealt{2021ApJ...920...94E}). As they contain coeval components, wide binaries have been proposed as alternative calibrators for gyrochronology \citep{2012ApJ...746..102C}. Useful systems for these purposes are binary pairs with orbits wide enough such that their components can be safely assumed to have evolved in isolation, and their separations on the sky are such that their stellar parameters, including their rotation periods, can be independently measured.  Specifically, the rotational evolution of their components must not have been affected by their distant companion (this requires pericenter distances between the components larger than $\sim 500$ au; \citealt{2022ApJ...930....7A}).
These systems are relatively abundant \citep[e.g.,][]{2004ApJ...601..289C, 2007AJ....133..889L, 2017MNRAS.472..675A, 2018MNRAS.480.4884E}. Pairs containing an evolved component can be used for calibration purposes, because the evolved star can be used to determine the age of the system, and the non-evolved MS companion can have a measurable rotation period (Godoy-Rivera et al. in preparation). However, the difficulty in obtaining rotation periods for specific sets of old stars in the field has not allowed this to succeed yet.  A complementary approach based on the coevality of wide binary members can be taken with MS-MS pairs, as they can be used to test gyrochronology relations if both stars have measured rotation periods \citep[e.g.,][]{2017ApJ...835...75J, 2022ApJ...930...36O, 2022ApJ...936..109P}. This latter approach is the focus of this work.

For the MS-MS wide binary test, there are two requirements: finding wide binaries, and determining the rotation periods of their components. Both tasks have become easier in recent years. For the binary search, many catalogs of pairs have been constructed, with fields-of-view encompassing the whole sky \citep[e.g.,][]{2017MNRAS.472..675A, 2018MNRAS.480.4884E, 2020ApJS..247...66H, 2021MNRAS.506.2269E} and some being focused specifically on the \textit{Kepler} field \citep[e.g.,][]{2017ApJ...835...75J,2018MNRAS.479.4440G}. For the rotation periods, the \textit{Kepler} and K2 missions \citep[][]{2010Sci...327..977B, 2014PASP..126..398H}, while being mainly focused on the search for Earth-like exoplanets, have opened the gate for works determining rotation periods for several tens of thousands stars with high precision \citep[e.g.,][]{2014ApJS..211...24M,2019ApJS..244...21S,2021ApJS..255...17S, 2020A&A...635A..43R, 2021ApJ...913...70G}. 

In order to put gyrochronology to the test, the present work focuses on assembling samples of stellar pairs with coeval components, for a range of ages, and with precise rotation periods independently measured for both components (see Section \ref{sec:data_periods}), and contrasts them against the predictions of current gyrochronology relations.  Coeval pairs are obtained both from wide binary catalogs and by pairing individual stars in clusters. 

This paper is structured as follows. In Section \ref{sec:rot_test} we introduce our new test designed to quantitatively examine the gyrochronology relations. Section \ref{sec:data} describes the sources of the data used, as well as the construction of the binary sample, the control samples of random pairs, and star cluster members. We show our results and discuss their implications in Section \ref{sec:discussion}. We discuss the evidence for a degradation of the current gyrochronology relations towards old ages in Section \ref{sec:degradation}. We conclude in Section \ref{sec:conclusions}.

\section{The Rotation Period Test} \label{sec:rot_test}

In this work, we compare measured rotation rates with three of the more recent gyrochronology relations available in the literature, namely, \citet{2015MNRAS.450.1787A}, \citet{2019AJ....158..173A}, and \citet{2020A&A...636A..76S}. Broadly speaking, the first two relations take the form of the rotation period being equal to a function of the stellar age times a function of the color. The third relation is a semi-theoretical model that predicts rotation periods given an age and mass, thus we use an interpolation grid to mimic a continuous function as in the first two relations. 

\begin{figure}
    \centering
    \includegraphics[width=0.95\columnwidth]{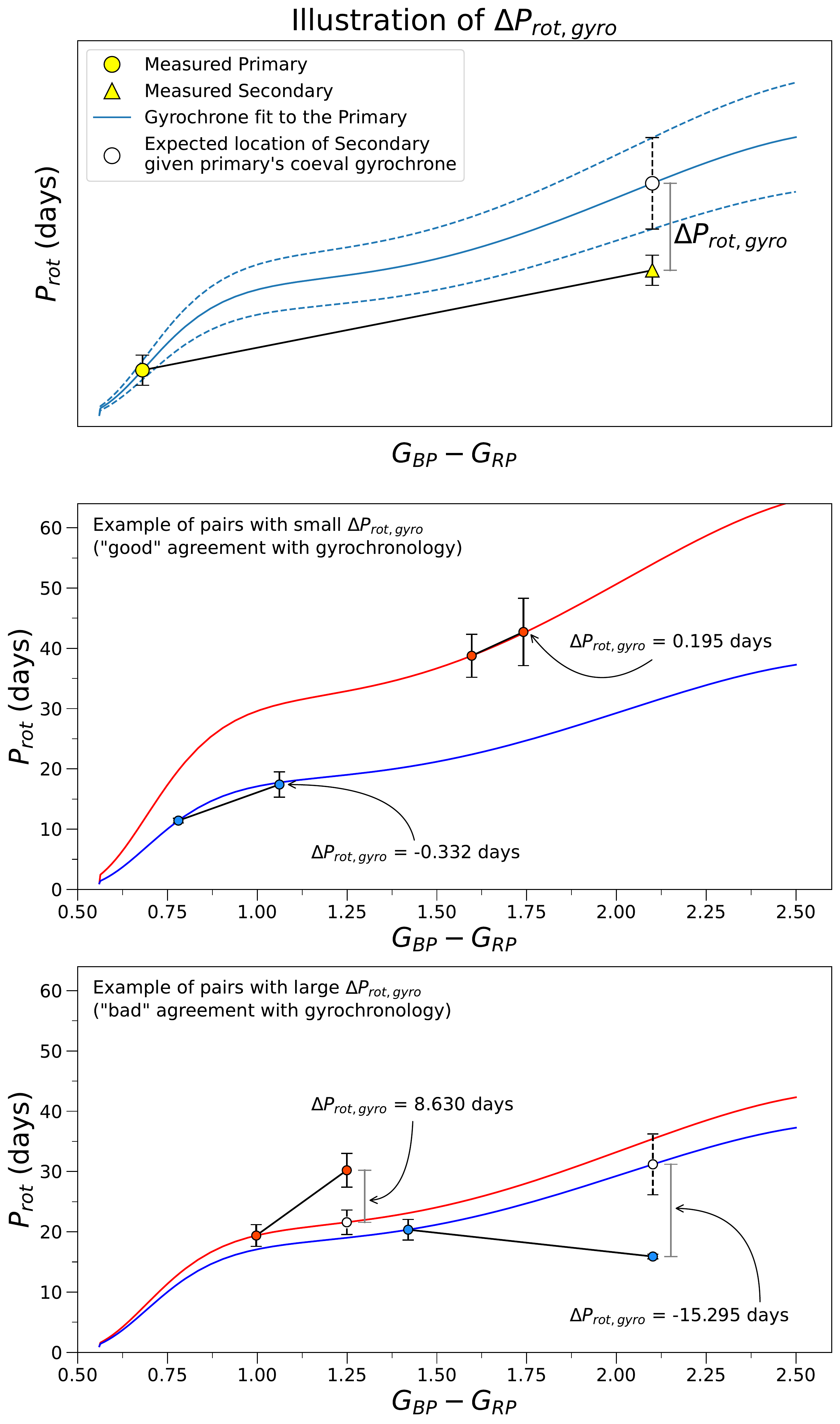}
    \caption{Top: schematic of the rotation period test and its error components. The yellow circle and triangle are the components of a representative binary, the blue line is the gyrochrone (using the \citealt{2019AJ....158..173A} relation) that, by construction, goes through the primary. $\Delta P_{rot, gyro}$ is, then, the difference between the measured and expected rotation periods of the secondary (Equation \ref{eq:gyro}). The latter corresponds to the rotation period of the primary's gyrochrone evaluated at the secondary's color. The error in the difference, $\sigma_{\Delta P_{rot, gyro}}$, is the sum in quadrature of the measurement error of the secondary's rotation period (shown as the secondary's error-bar) and the propagated error of its expected rotation period (shown as the dashed error-bar, calculated from the dashed gyrochrones corresponding to the primary's rotation period plus/minus its measurement error). Middle: example of two binary candidates with small $\Delta P_{rot,gyro}$ values; the red pair is KIC 8676609/KIC 8676604 and the blue pair is KIC 6678383/KIC 6678367. Bottom: example of two binary candidates with large $\Delta P_{rot,gyro}$ values; the red pair is KIC 11861595/KIC 11861593 and the blue pair is KIC 11600772/KIC 11600744. A smaller $\Delta P_{rot,gyro}$ means a better agreement with the gyrochronology relations, while a larger one means a worse agreement.}
    \label{fig:rot_eg}
\end{figure}

In order to quantify the agreement of our samples with these relations, we designed the following test, based on the assumption that the components of wide binaries are coeval. For any given pair suitable to be placed on the color vs. rotation period diagram (i.e., where we have period and color measured for both components independently), our test calculated the quantity $\Delta P_{rot, gyro}$. We illustrate this in the top panel of Figure \ref{fig:rot_eg} and describe it as follows:

\begin{enumerate}
    \item Given a wide binary pair and a gyrochronology relation, the age of the primary star was calculated by solving the gyrochronology equation that goes through its measured rotation period and color (yellow circle), effectively returning the gyrochrone that contains it (solid blue line). We chose to fit the primary because brighter stars tend to have better measured rotation periods, thus (presumably) yielding a better age estimation.
    
    \item Under the assumption that the components of the pair are coeval, the age of the primary was then assumed for the secondary. With this, we determine an expected rotation period for the secondary ($P_{rot, expected}$) given its color (i.e., as predicted by the gyrochronology relation being tested; open black circle).
    
    \item The difference between the measured rotation period ($P_{rot,measured}$; yellow triangle) and the expected rotation period for the secondary was calculated as a proxy for how good the agreement is between the data and the gyrochronology prediction. In other words, we define:
    \begin{equation}
      \Delta P_{rot, gyro} = P_{rot,measured} - P_{rot,expected}. \label{eq:gyro}
    \end{equation}
\end{enumerate}

For the coeval components of a wide binary that are in good agreement with the gyrochronology relations, the $\Delta P_{rot, gyro}$ value should be relatively small, as illustrated in the middle panel of Figure \ref{fig:rot_eg}. On the other hand, a pair with components in bad agreement with gyrochronology (or a random pair with non-coeval, unrelated components) should yield a large $\Delta P_{rot, gyro}$, as illustrated in the bottom panel in Figure \ref{fig:rot_eg}. 

One important consideration, however, is that $\Delta P_{rot, gyro}$ may not be directly comparable among different pairs and samples, as our test depends essentially on the measured rotation periods for both components, and their errors. As later discussed in Section \ref{sec:discussion}, the rotation periods and their errors distribution vary considerably among different samples. With this as a motivation, we calculated the error on $\Delta P_{rot,gyro}$($=\sigma_{\Delta P_{rot, gyro}}$) as the sum in quadrature of the measurement error of the secondary's rotation period, and the error of its expected rotation period. The latter is calculated by propagating the error of the primary's rotation period through its gyrochrone (dashed error-bar in Figure \ref{fig:rot_eg}). In the case of \citet{2015MNRAS.450.1787A} and \citet{2019AJ....158..173A}, the models include uncertainties to their fit parameters, and thus we also propagate them through their gyrochrones, contributing to the expected rotation period error. We could not do the same with the \citet{2020A&A...636A..76S} model, as it is not an analytical relation with parameter uncertainties. With this error, a new parameter can be defined as:

\begin{equation}
    x = \left|\frac{\Delta P_{rot, gyro}}{\sigma_{\Delta P_{rot, gyro}}}\right|. \label{eq:x}
\end{equation}

This x-parameter then quantifies the significance of a given $\Delta P_{rot,gyro}$ value by dividing it by its error. If $x\leq1$, then our pair is in very good agreement with the corresponding gyrochronology model prediction, as $\Delta P_{rot, gyro}$ is smaller than its propagated measurement error. We defined $x$ as an absolute value for simplicity, because, as will be shown in Section \ref{sec:discussion}, the distributions of $\Delta P_{rot,gyro}$ tend to be symmetric. Further, this definition of the x-parameter makes the test essentially blind as to whether the calculation of $\Delta P_{rot, gyro}$ is done with respect to the primary or the secondary of the given pair.\footnote{This paper is accompanied by Python functions that allow the user to calculate $\Delta P_{rot,gyro}$, $\sigma_{\Delta P_{rot,gyro}}$, and the x-parameter, given the input data (rotation periods and colors) for the components of a given binary star (or any pair of stars), as well as a choice of a gyrochronology relation to use: \url{https://github.com/jsilvabeyer/wbgyro23}}.

\section{Data}\label{sec:data}

Here we present the data search, cross-matching, and filtering process that yielded our tested samples. Due to the specific range of applicability of each of the relations under scrutiny \citep{2015MNRAS.450.1787A,2019AJ....158..173A,2020A&A...636A..76S}, we had to construct slightly different samples for each model, which we detail below.

\subsection{Cluster members} \label{sec:data_clusters}

Empirically-determined timescales for star formation in giant molecular clouds are around 4-6 Myr \citep{2022MNRAS.516.4612T}.  One can thus consider that stars in clusters older than, say, 10 times such timescales, should be well approximated as being coeval among themselves, as any age dispersion due to their formation epoch would not be larger than 10\% of their current age.  This of course is the reason why clusters are commonly used as calibrators for gyrochronology relations.  For the purposes of the present work, by pairing stars in clusters older than 100 Myr we would be assembling pairs of coeval stars, which is a first step for our test.  Moreover, if gyrochronology works, and current relations correctly capture that property, we would expect that pairs of stars belonging to the same cluster will show, by construction, a good agreement with gyrochronology.

As introduced in Section \ref{sec:rot_test}, our rotation period test takes pairs of stars as input. In order to apply our technique for a given star cluster, we randomly paired its members, so that these pairs retain the coevality inherited by belonging to the same (sufficiently old) cluster. Four clusters were used to form this sample: Praesepe (age $\sim$ 650 Myr) and NGC6811 (age $\sim$ 1 Gyr) from \citet{2021ApJS..257...46G}; Ruprecht 147 (age $\sim$ 2.5 Gyr) from \citet{2020A&A...644A..16G} and \citet{2020ApJ...904..140C}; and NGC6819 (age $\sim$ 2.5 Gyr) from \citet{2015Natur.517..589M}. We highlight the wide range of ages spanned by our cluster sample, as this is central to the gyrochronology examinations we perform below.

We compiled rotation period uncertainties from different sources. For Praesepe, as the original source \citep{2017ApJ...839...92R} did not provide rotation period errors, we adopted the median fractional error from the other sources ($\sigma_{P_{rot}}/P_{rot}\approx7\%$) as a representative value. For NGC6811 we compiled the corresponding rotation period uncertainties from the original sources \citep[]{2019ApJS..244...21S, 2021ApJS..255...17S, 2019ApJ...879...49C, 2011ApJ...733L...9M}. For the Ruprecht 147 stars from \citet{2020ApJ...904..140C}, no period uncertainties were reported either, so we followed the same procedure as for Praesepe. Finally, the NGC6819 stars from \citet{2015Natur.517..589M} and the Ruprecht 147 stars from \citet{2020A&A...644A..16G} had period errors already reported, and so we used them.

Note that for Praesepe and NGC6811 we took the stars that were astrometrically classified as probable cluster members by \citet{2021ApJS..257...46G}. In other words, these stars are fully consistent with being members of the clusters, but at the same time, rotational outliers are present in the sample (see figures 9 and 10 in \citealt{2021ApJS..257...46G}). Similarly, for the \citet{2020A&A...644A..16G} Ruprecht 147 sample, we discarded stars with low confidence in the rotation period measurement. These choices were made in order to perform meaningful comparisons with field stars, and the implications of the above on the gyrochronology tests are discussed in Section \ref{sec:discussion}.

For photometric colors, we performed a cross-match between the cluster sample (as well as all the other samples described in the rest of this section), and the Gaia DR2 Catalog \citep{2018A&A...616A...1G}, to use the $G_{BP}-G_{RP}$ color as a proxy for mass. For consistency, we defined the primary for all our samples as the brighter star in the pair (i.e., the one with lower apparent $G$ magnitude). We de-reddened the Gaia photometry of the clusters by adopting a global $A_V$ extinction for each one, then computing $A_G$ and $E(\text{BP}-\text{RP})$ following the approach of \citet{2021ApJS..257...46G} on a star-by-star basis (where the extinction coefficients depend on both $A_V$ extinction and $G_{BP}-G_{RP}$ color). A summary of the global V-band extinction values adopted and their sources, along with mean values for Gaia extinction $A_G$ and reddening $E(\text{BP}-\text{RP})$ for each cluster, is presented in Table \ref{tab:ext}.

\begin{table}
    \centering
    \begin{tabular}{l|l|l|l|c}
         \hline
         Cluster  & $A_V$  & Mean $A_G$ & Mean $E(\text{BP}-\text{RP})$ & Source \\
            &  (mag) & (mag)        & (mag) & \\ \hline
         Praesepe & 0    & 0 &  0  &  1    \\
         NGC6811  & 0.147   &  0.119 & 0.068 & 2\\
         Ruprecht 147 & 0.3 &  0.235   &    0.136 & 3   \\
         NGC6819 & 0.465 & 0.382       &  0.217  & 4   \\ \hline
    \end{tabular}
    \caption{Reddening and extinction parameters for the clusters samples. Sources: (1) \citet{2019AJ....158..173A}; (2) \citet{2021ApJS..257...46G}; (3) \citet{2020ApJ...904..140C}; (4) \citet{2015Natur.517..589M}. The mean Gaia extinctions and reddening parameters we have calculated show an approximate linear relation of $A_G \sim 1.7 \times E(\text{BP}-\text{RP})$, with a slightly shallower slope than the one computed by \citet{2018A&A...616A...8A}.}
    \label{tab:ext}
\end{table}

\begin{figure}
    \centering
    \includegraphics[width=\columnwidth]{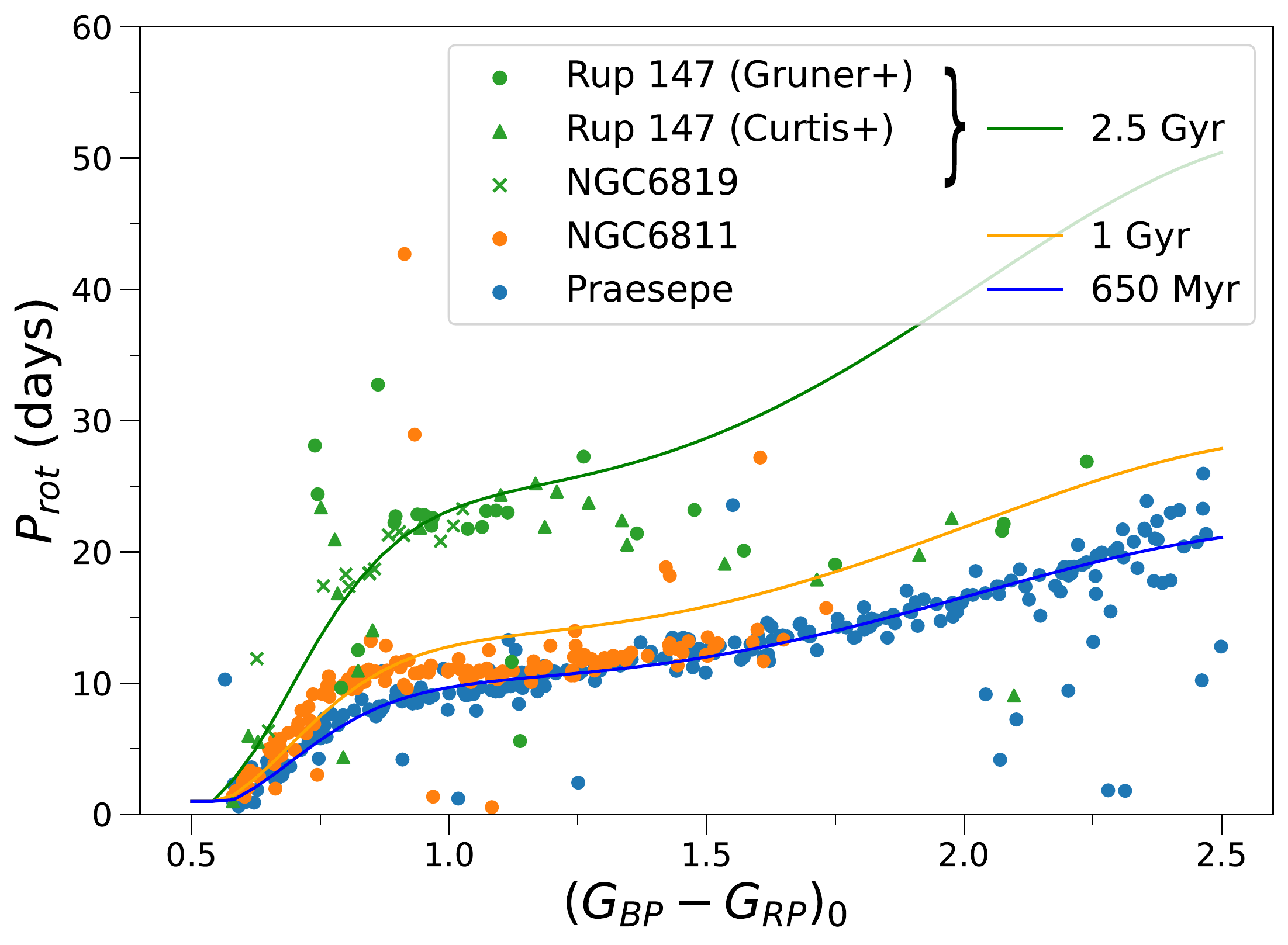}
    \caption{Color vs. rotation period for the four open clusters used in this work. Blue dots are for members of the Praesepe cluster, orange dots are for the NGC6811 cluster, green dots are the Ruprecht 147 stars from \citet{2020A&A...644A..16G}, green triangles are the Ruprecht 147 stars  from \citet{2020ApJ...904..140C}, and green crosses are for the NGC6819 cluster members. The solid lines are gyrochrones from the \citet{2019AJ....158..173A} relation, adopting the age of each cluster. Note that we combined the $\sim 2.5$ Gyr clusters, Ruprecht 147 and NGC6819. See Section \ref{sec:data_clusters} for details and discussion.}
    \label{fig:clusters}
\end{figure}

Since all gyrochronology relations have a minimum color for which an age can be measured (colors bluer than this step onto the radiate-envelope regime), we only kept stars with\footnote{Being $(G_{BP} - G_{RP})_0 = (G_{BP} - G_{RP}) - E(\text{BP} - \text{RP})$ the color index corrected by reddening.}$(G_{BP} - G_{RP})_0 > 0.562$. Additionally, we ignored fully convective stars -- $(G_{BP} - G_{RP})_0 > 2.5$ --, as they tend to break apart from the usual $\sqrt{t}$ dependence, and in this regime the calibrations are considerably worse (see \citealt{2019AJ....158..173A}; \citealt{2022ApJ...936..109P}; \citealt{2022AJ....164..251L}). The color-cuts described above were applied directly to our data sample, i.e., on the Gaia DR2 colors. This is also the color-index in which the \citet{2019AJ....158..173A} was calibrated.

To implement these color-cuts for the \citet{2015MNRAS.450.1787A} and \citet{2020A&A...636A..76S} models, however, a few extra steps were required due to the specifics of each relation. The \citet{2015MNRAS.450.1787A} relation was calibrated with the $(B-V)_0$ color, so we transformed the stars' Gaia DR2 $(G_{BP}-G_{RP})_0$ into $(B-V)_0$ using a 1 Gyr PARSEC isochrone \citep[][]{2012MNRAS.427..127B}. We used the PARSEC models as they are empirically calibrated, and because the alternative analytical photometric relationships we tested \citep{2010A&A...523A..48J, 2018gdr2.reptE....V} yielded inconsistent results in the relevant color range. We also needed to impose a slightly higher minimum color for the pair components, as the \citet{2015MNRAS.450.1787A} relation has a lower color-boundary of $(B-V)_0 = 0.45$, which corresponds to $(G_{BP} - G_{RP})_0 \approx 0.582$. On the other hand, the \citet{2020A&A...636A..76S} relation was calibrated with mass instead of color, so we used the same PARSEC isochrone to convert the stars $(G_{BP}-G_{RP})_0$ colors into mass. Moreover, we needed to impose stricter constraints to the colors and rotation periods of the stars. First, this model only reported gyrochrones for masses between $0.45 M_{\odot}$ and $1.25 M_{\odot}$, so we had to impose the equivalent filter for the Gaia color ($0.592<(G_{BP} - G_{RP})_0<2.346$). Second, the \citet{2020A&A...636A..76S} model only spans the ages between $0.1$ Gyr and $4.57$ Gyr, thus, we only kept stars for which their rotation periods (plus/minus their uncertainties) were between these two limiting gyrochrones.

To ensure that we are using well measured rotation periods, we required stars to have $P_{rot} > 3 \sigma_{P_{rot}}$. Additionally, in order to filter out potential close binaries and evolved stars from our sample, we imposed two further constraints to our stars. First, for each cluster we downloaded a PARSEC isochrone of its respective age, and eliminated all stars with absolute magnitudes more luminous than $0.38$ mag from the isochrone. This limit is calculated as $\approx 1/2 \times 2.5 \log_{10}(2)$, and corresponds to half of the magnitude difference between the isochrones of single stars and equal-mass photometric binaries (i.e., systems with twice the flux).  We did this to identify and discard stars that may be part of close, unresolved binaries. \citet{2020A&A...644A..16G} used a similar constraint for their Ruprecht 147 sample, thus, we also discard the stars they flagged as photometric binaries. Second, we dropped all stars more luminous than an absolute magnitude of $2.85$, as they probably include evolved stars.

Finally, in order to add robustness to our random pairing method, and increase the statistics of our sample, we repeated the process of randomly pairing the members of any given cluster a total of 20 times. This ensured a Cluster sample 20 times larger than the original, while simultaneously keeping the distributions of the stars' properties (color, rotation period, etc.) identical, as each member was over-sampled the same number of times. 

After our quality cuts, regarding the \cite{2019AJ....158..173A} relation, we were able to run our test for 125 pairs (2500 pairs after the over-sampling process) for Praesepe, 75 (1500) pairs for NGC6811, 24 (480) pairs for Ruprecht 147, and 7 (140) pairs for NGC6819. Following \citet{2020ApJ...904..140C}, we combined the Ruprecht 147 and NGC6819 samples because of their practically identical ages, thus constructing a sample of 31 (620) pairs composed by $\sim 2.5$ Gyr-old cluster members. For the \citet{2015MNRAS.450.1787A} relation, the analogous numbers were: 124 (2480 after oversampling) pairs for Praesepe, 74 (1480) pairs for NGC6811, 24 (480) pairs for Ruprecht 147, and 7 (140) pairs for NGC6819. For the \citet{2020A&A...636A..76S} relation, the stricter constraints reduced the sample sizes, resulting in the final number of: 105 (2100) pairs for Praesepe, 65 (1300) pairs for NGC6811, 20 (400) pairs for Ruprecht 147, and 7 (140) pairs for NGC6819.

In Figure \ref{fig:clusters} we show the rotational sequences of the four clusters used (after applying our quality cuts). For each one, we also plot a gyrochrone (from \citealt{2019AJ....158..173A}) of the corresponding cluster age, which are shown in the label for reference. As \citet{2019AJ....158..173A} based their relation on Praesepe data, the 650 Myr line fits almost perfectly our Praesepe sample, by construction. 

On the other hand, faster-than-expected rotational sequences can be seen for the 1 Gyr cluster NGC6811, and the 2.5 Gyr clusters NGC6819 and Ruprecht 147. More extensive studies of the NGC6811 cluster have revealed the apparent merging of its rotational sequence (for colors redder than $G_{BP}-G_{RP}\gtrsim 1$) with the Praesepe cluster (\citealt{2019ApJ...879...49C}; see also \citealt{2011ApJ...733L...9M} and \citealt{2018ApJ...862...33A}). This is discussed further in Section \ref{sec:discussion_clusters}.

\begin{figure*}
    \centering
    \includegraphics[width=1.9\columnwidth]{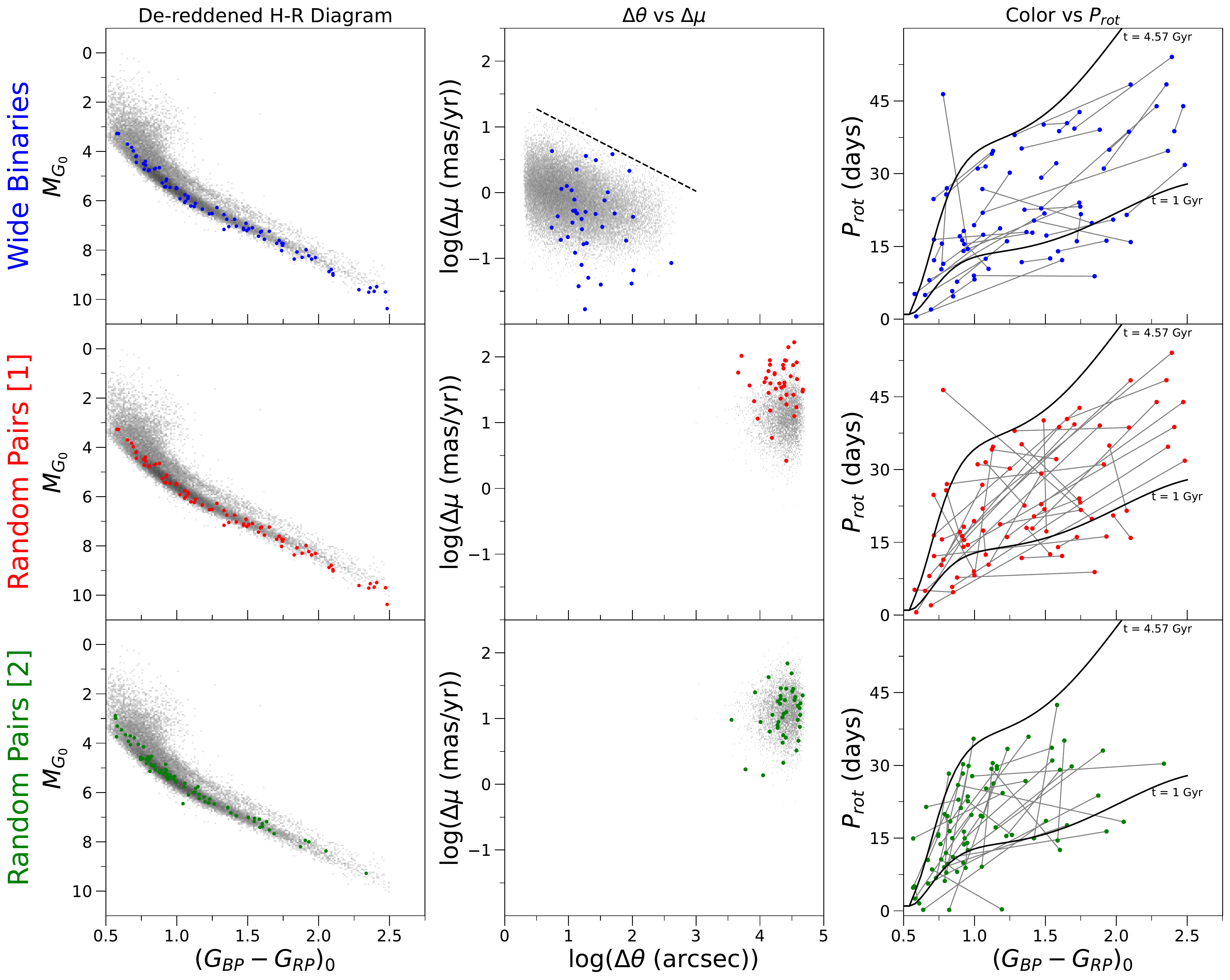}
    \caption{Grid of plots characterizing the Wide Binaries, Random Pairs [1], and Random Pairs [2] samples. The panels on the left column are HRDs, the ones in the centre column are angular separation vs. proper motion difference, and the right column are color vs. rotation period plots, connecting the pairs with gray lines (for reference, we plot \citealt{2019AJ....158..173A} gyrochrones for 1 Gyr and the age of the Sun found by \citealt{2012Sci...338..651C}, 4.567 Gyr). The top row shows plots for the Wide Binaries sample (blue), the middle for Random Pairs [1] (red), and the bottom for Random Pairs [2] (green). For both samples of random pairs, we only show a subset of 40 pairs (to match the number of wide binaries shown).
    The gray background on the HRDs are the stars from the \citet{2014ApJS..211...24M} catalog. The gray background for the middle plots are pairs from \citet{2018MNRAS.480.4884E} for the first row, and the whole sample of Random Pairs [2] for the second and third rows. In the centre panel for the Wide Binaries we plot a dashed line with slope of $-1/2$; pairs above this are considered to be suspect due to their large proper motion differences.}
    \label{fig:3x3}
\end{figure*}

\subsection{Wide Binaries}

We compiled a sample of \textit{Kepler}-field wide binaries, given the availability of literature rotation periods for stars in that field (see Section \ref{sec:data_periods}). We do this by combining existing catalogs from dedicated searches. Our binary sample was composed of four sources: \cite{2016MNRAS.455.4212D}, \cite{2017ApJ...835...75J}, \cite{2018MNRAS.479.4440G}, and \cite{2018MNRAS.480.4884E}.

We de-reddened the Gaia colors following the same procedure as in Section \ref{sec:data_clusters}. In this case, we queried the \texttt{Bayestar19} map \citep{2019ApJ...887...93G} for extinction values for each star using the \texttt{GALExtin} website \citep{2021MNRAS.508.1788A}. We then calculated the absolute magnitude for each star using the distances computed by \citet{2018AJ....156...58B}. 

The first row of plots in Figure \ref{fig:3x3} characterizes the Wide Binaries sample. The top-left plot is an HRD and the top-centre plot shows angular separation vs. proper motion difference. For guidance, in the background of these panels we respectively show the \citet{2014ApJS..211...24M} stars compared to the Kepler-field HRD, and the \citet{2018MNRAS.480.4884E} binary pairs to illustrate their astrometric selection. The HRD shows that, by construction, most of the binaries components are MS stars, making them suitable for our gyrochronology tests. The centre plot shows that most pairs of the Wide Binaries sample fall between $\sim$ 5\arcsec \ and 500\arcsec \ of separation. 
These parameters are consistent with gravitationally bound systems, as indicated by the background pairs. On the centre plot, we draw a dashed line (with a $-1/2$ slope corresponding to a Keplerian orbit) to identify pairs that could be considered questionable due to their large proper motion differences \citep[e.g., see][]{2017MNRAS.472..675A, 2018MNRAS.480.4884E, 2019MNRAS.490.2448R}. For our combined binary sample, we discard any pairs above this line, as they are probably chance alignments or one of their components is a close, unresolved binary. We do this to compensate for possible heterogeneities in the selection in the aforementioned source binary catalogs.

\subsection{Rotation periods} \label{sec:data_periods}

Stars in the \textit{Kepler} field are arguably the best sample to use for our test, as the high precision photometry and observing baseline that encompasses many years translate into reliable rotation periods. In order to have a relative homogeneity of rotation period sources, we used primarily rotation periods from \cite{2019ApJS..244...21S}, and adopted the values from \cite{2014ApJS..211...24M} in cases where stars were not found in the former. The wide binary lists from \citet{2017ApJ...835...75J} and \citet{2018MNRAS.479.4440G} compiled rotation periods too, so we used their values for stars where the two primary catalogs did not report rotation periods. \footnote{\cite{2016MNRAS.455.4212D} also compiled rotation periods, but they were exclusively from the \cite{2014ApJS..211...24M} catalog, so no extra rotation periods were provided.} 

While all these periods ultimately come from \textit{Kepler} observations, the different  sources calculate them using a variety of methods (e.g., Lomb-Scargle periodograms, wavelet analysis, auto-correlation function). Thus, we may suspect the rotation periods to have different biases and selection effects. To quantify this, we used the Wide Binaries and NGC6811 stars with more than one period source, and compared the rotation periods coming from the different references. We found that more than 98\% of these stars had rotation periods that agree at the $95\%$ level or higher. This excellent agreement validates the reliability of the literature rotation periods, and we thus consider the prioritization of specific period sources to have a negligible effect on our work.
\footnote{The one caveat to this relates to the rotation period errors reported by  \cite{2014ApJS..211...24M}, but we further discuss this in Section \ref{sec:discussion_random_pairs}.}

After applying the same color, $P_{rot}/\sigma_{P_{rot}}$, and HRD filters used for the clusters (see Section \ref{sec:data_clusters}; in this case using a 2 Gyr isochrone), we were able to run our test for a total of 40 Wide Binaries for \cite{2019AJ....158..173A}, 39 for \cite{2015MNRAS.450.1787A}, and 19 for \cite{2020A&A...636A..76S}. We show the former of these samples in the color vs. rotation period diagram on the top-right panel of Figure \ref{fig:3x3}. Note that the smaller size of our \cite{2020A&A...636A..76S} Wide Binaries sample is due to the stricter constraints we had to apply due to the more restricted regime of applicability of their relation (see Section \ref{sec:data_clusters}).

\begin{figure*}
    \centering
    \includegraphics[width=0.95\textwidth]{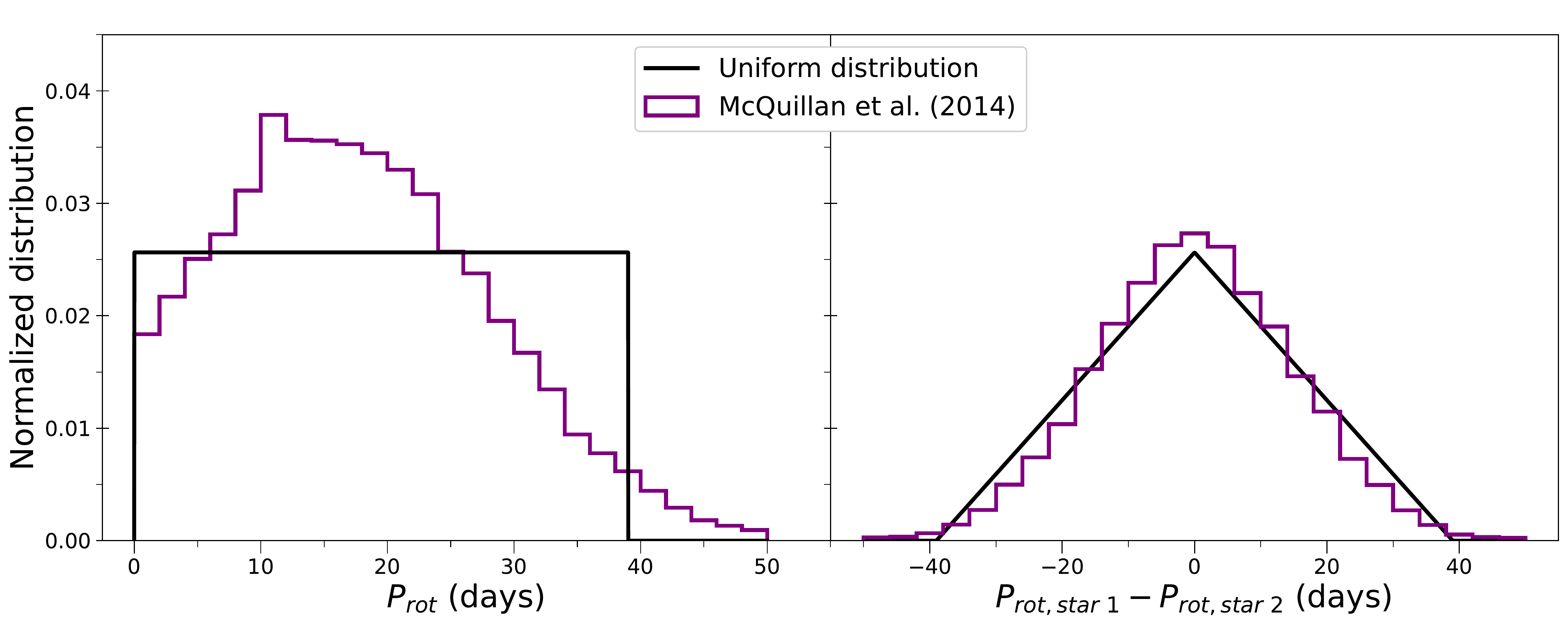}
    \caption{Left: distributions of rotation periods from the \citet{2014ApJS..211...24M} catalog (purple) and a uniform distribution between 0 and the 2$\sigma$ percentile ($\sim39$ days) of the \citet{2014ApJS..211...24M} catalog (black). Right: we subtracted the rotation periods of two randomly selected stars from the distributions on the left. We plot in purple the resulting difference distribution for \citet{2014ApJS..211...24M}, and in black the analytical PDF resulting from this random subtraction for the uniform distribution. Given its definition (Equation \ref{eq:gyro}), this sets our expectations for symmetric and centered-at-zero distributions for $\Delta P_{rot,gyro}$ in the tests that follow.}
    \label{fig:triang}
\end{figure*}

\subsection{Random Pairs} \label{sec:random_pairs}

Constructing a random pairs sample is important, as it gives us a statistical control group composed of stars that are not coeval. These samples are typically used in binary studies \citep[e.g.,][]{2007AJ....133..889L, 2019ApJ...871...42A, 2021ApJ...920...94E}. For this purpose, two different random pair samples were constructed.

\begin{itemize}
    \item Random Pairs [1]: random pairs generated from the components of the wide binaries sample. We took all the components of the wide binaries selected for the rotation period test (for their respective gyrochronology model) and paired them randomly, ensuring that all real pairs from the original sample were destroyed. 
    This guaranteed us a sample of pure random pairs with no real binaries in it. It was also generated so the stars in this sample of random pairs have, by construction, identical distributions of magnitudes, colors, $P_{rot}$, and $\sigma_{P_{rot}}$ as the wide binaries (because they are the same stars). This avoids potential systematic biases that could appear if we compare two samples with different distributions of stellar properties. As with the Clusters sample, we repeated the pairing process 20 times, over-sampling each star an equal amount of times, so the underlying distributions are kept identical.

    \item Random Pairs [2]: random pairs from \cite{2014ApJS..211...24M} stars. The \cite{2014ApJS..211...24M} catalog contains rotation periods for over 34,000 stars in the \textit{Kepler} field. We paired randomly every star in the catalog, constructing a sample of over 17,000 random pairs, with no repeated stars. We also de-reddened their colors through the same procedure as for the Wide Binaries. The result, after applying the previously mentioned color, $\sigma_{P_{rot}}$ and HRD constraints, was a sample of 11,011 random pairs. 
    This approach has the advantage of generating a very large number of pairs, making the sample significantly less sensitive to statistical variations, and avoiding the need for over-sampling a small population of stars. A drawback of this sample, however, is that it does not follow the same distribution of stellar properties as the Wide Binaries and Random Pairs [1]. This is reflected in the color and period distributions shown in the bottom-left and bottom-right panels of Figure \ref{fig:3x3}. Another aspect to consider for this sample, is that by being constructed exclusively from the \citet{2014ApJS..211...24M} catalog, it inherits its $\sigma_{ P_{rot}}$ values. We find that these tend to be significantly smaller than the $\sigma_{ P_{rot}}$ values of the other sources, which has important effects on the x-parameter we have previously defined. We discuss this effect in Section \ref{sec:discussion}.
    
\end{itemize}

With this, we have constructed two samples of random pairs with pros and cons that complement each other. The middle and bottom rows of Figure \ref{fig:3x3} show plots for the Random Pairs [1] (red points) and [2] (green points) samples, respectively. The centre column plots show that the random pairs have much wider separations and larger proper motion differences than the Wide Binaries, both of which are to be expected. Note that, as Wide Binaries and Random Pairs [1] are composed of the same stars (only the pairings are different), the HRD and color vs. rotation period plots for the upper and middle rows are virtually identical, except for the connecting lines in the latter panel.

\section{Results and Discussion}
\label{sec:discussion}

We now present and discuss the main results of our work. We ran our rotation period test in the four samples (Clusters, Wide Binaries, Random Pairs [1], Random Pairs [2]) and calculated $\Delta P_{rot,gyro}$ and $x$ for each of the pairs within them. As both \citet{2015MNRAS.450.1787A} and \citet{2019AJ....158..173A} yielded very similar results, we only discuss the latter, being the more recent work. Likewise, the results of the \citet{2020A&A...636A..76S} relation are similar to the other relations, with some specific differences that we discuss in Section \ref{sec:degradation}.

For conciseness, this section illustrates the figures obtained by running the rotation period test for the \citet{2019AJ....158..173A} relation, and summarizes the results for both \citet{2019AJ....158..173A} and \citet{2020A&A...636A..76S} in Table \ref{tab:results}. We report all the data used in this paper in Appendix \ref{sec:data_table}.

\subsection{Expectations: comparing with distributions of rotation period differences} \label{sec:results_expectations}

Since $\Delta P_{rot,gyro}$ is the difference of two variables with relatively similar distributions (i.e., $\Delta P_{rot,gyro} = P_{rot,measured} - P_{rot,expected}$ for the secondary of a given pair), we set the context of our expectations by exploring the distribution that results from subtracting two random variables that follow the same distribution. In the left panel of Figure \ref{fig:triang}, we plot (in black) a uniform distribution of rotation periods, and (in purple) the distribution of the rotation periods from \citet{2014ApJS..211...24M}.\footnote{In order to keep consistency with the data, the rotation periods used are those of the components of the Random Pairs [2] sample.} In the right panel, we plot (in black) the resulting distribution obtained by subtracting two random variables that followed the uniform distribution, and (in purple) the one obtained by subtracting two random variables that followed the \citet{2014ApJS..211...24M} distribution.

The black distribution on the right-hand panel has a triangular shape, it is symmetric and centered at zero, and it represents the floor for a $\Delta P_{rot}$ distribution, i.e., the case where all pairs of stars have unrelated components, and thus their rotation period values should be completely independent (the components of some random pairs may have similar ages, and thus small $\Delta P_{rot}$, but this happens by chance, not systematically). It is not a uniform distribution, despite the distributions of the quantities in the difference being uniform themselves, and shows a peak at zero. This means that, even among pairs of unrelated, non-associated stars, a preference for {\it seemingly correlated} rotation periods (and therefore ages) is to be expected. 

The purple distribution is very similar to the black one, although slightly taller and narrower (somewhat resembling a more Gaussian shape). This demonstrates that we should expect the $\Delta P_{rot,gyro}$ distribution of each of our samples (even the random pairs) to be centered on zero and fairly symmetric. This also justifies our assumption of taking the absolute value in the definition of the x-parameter (Equation \ref{eq:x}).

\subsection{Clusters} \label{sec:discussion_clusters}

\begin{figure*}
    \centering
    \includegraphics[width=2.1\columnwidth]{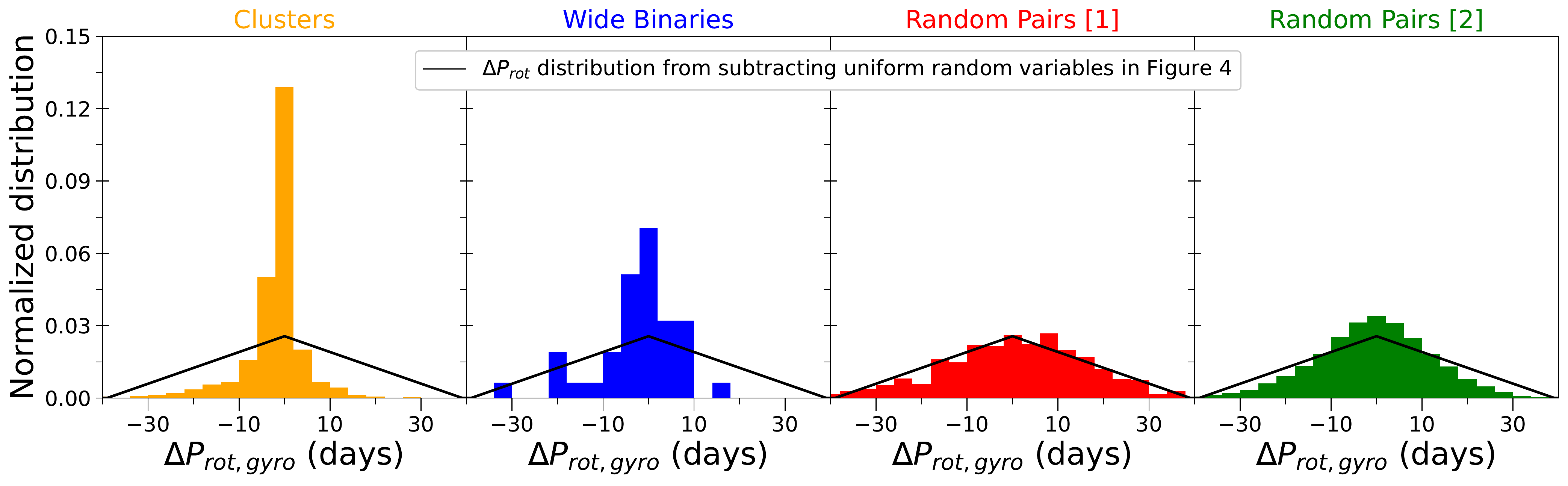}
    \caption{Normalized distributions for the  $\Delta P_{rot, gyro}$ values for the \citet{2019AJ....158..173A} relation for the combined Clusters (orange), Wide Binaries (blue), Random Pairs [1] (red), and Random Pairs [2] (green) samples. We also plot, as a black line, the triangle distribution from the right panel of Figure \ref{fig:triang}, for comparison purposes. All the distributions are centered around zero, but those of the Wide Binaries and Clusters are narrower and have a higher peak than the Random Pair ones, implying a better agreement with gyrochronology. The Random Pairs [1] and [2] distributions appear similar to the triangle distribution, which is in agreement with our expectations (see Section \ref{sec:results_expectations}).}
    \label{fig:prot_check_hist}
\end{figure*}

In Figure \ref{fig:prot_check_hist}, we show the $\Delta P_{rot,gyro}$ distributions resulting from applying our test with the \citet{2019AJ....158..173A} relation to the four samples constructed in Section \ref{sec:data}.

As it can be seen, the Clusters sample produces the best results among our four samples, with its $\Delta P_{rot, gyro}$ distribution displaying the highest peak and the narrowest width of all distributions, i.e., a better statistical agreement with gyrochronology.  Of course, this is the expected outcome, given that the gyrochronology relations have been calibrated using the color vs. rotation period distributions measured for stars in clusters.  Indeed, \citet{2019AJ....158..173A} used the Sun and the Praesepe cluster to constraint their relation. 

In Table \ref{tab:results} we present the fraction of pairs that satisfy $x\leq1,2,3$. As can be inferred from Equation (\ref{eq:x}), this means the fraction of pairs for which $\Delta P_{rot, gyro}$ falls inside $1,2,3$ times its uncertainty. Here, the fraction of pairs with $x$ smaller than a given value is always higher for the Clusters sample than for the other samples, further demonstrating their better agreement with gyrochronology.

\begin{figure}
    \centering
    \includegraphics[width=\columnwidth]{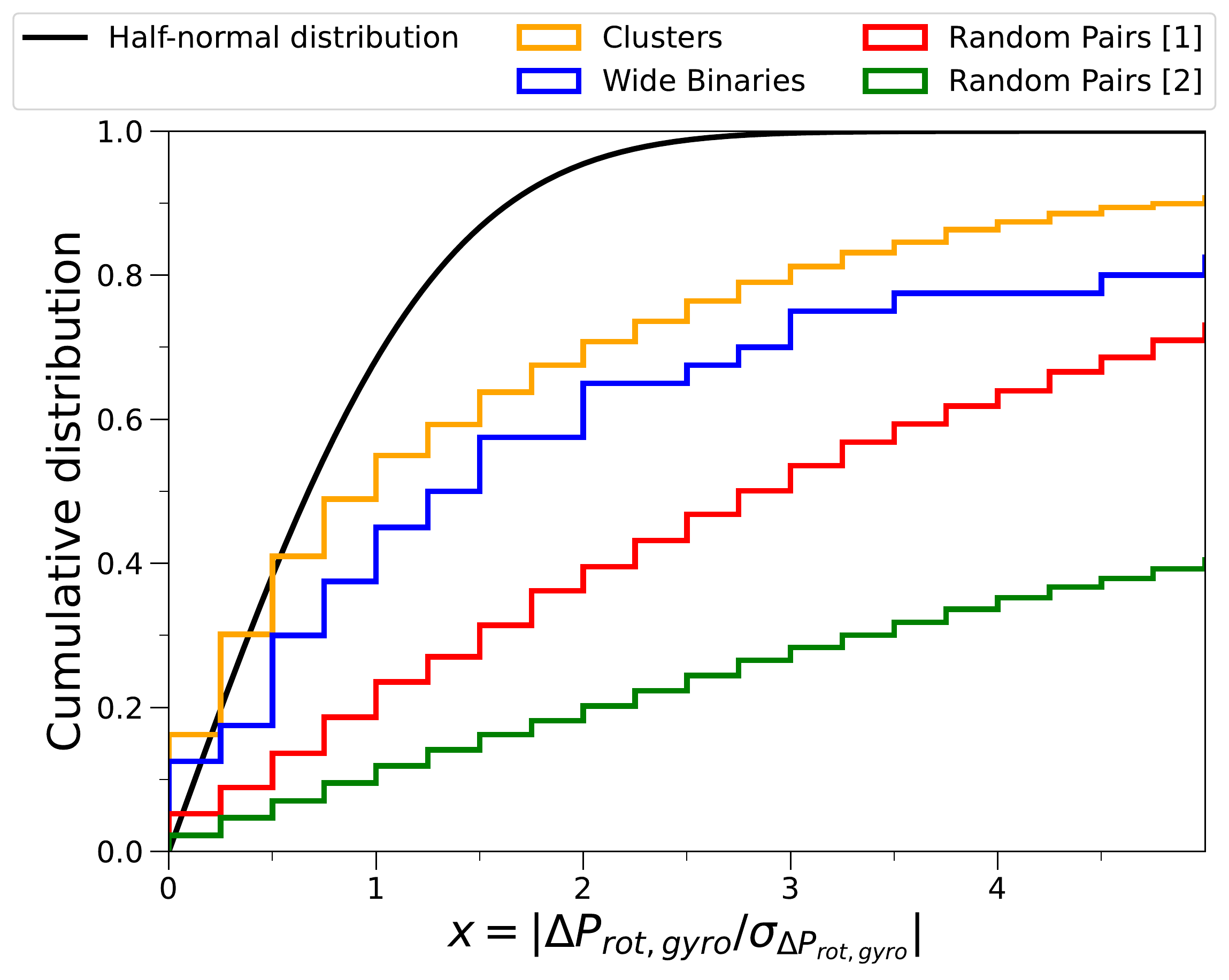}
    \caption{Cumulative distributions of $x$ (see Equation \ref{eq:x}), for the tests performed with the \citet{2019AJ....158..173A} relation. As the histograms are normalized, the cumulative probability is equivalent to the values in Table \ref{tab:results} (e.g. the $x\leq1$ columns in the Table are exactly equal to the $x=1$ points in this figure). The black continuous line is the cumulative distribution function (CDF) of the half-normal distribution, which we plot for comparison purposes. While the Cluster distribution is below the half-normal distribution, it raises significantly more rapidly than the other samples. The Wide Binaries distribution is also clearly higher than the random pairs. This indicates that the samples of pairs with coeval components present a better agreement with gyrochronology than the random pairs samples.}
    \label{fig:hist_x} 
\end{figure}

In Figure \ref{fig:hist_x}, we present the comparison between the cumulative distributions of the x-parameter for the samples. Thus, the cumulative probability where $x=1,2,3$, corresponds to the values reported in Table \ref{tab:results}. The distribution of the Clusters climbs much more rapidly in comparison with the other distributions, reflecting its better concordance (i.e., a narrower peak in Figure \ref{fig:prot_check_hist}) with the gyrochronology relations by construction.

In the hypothetical case that the agreement of a given sample of pairs with a gyrochronology relation were ``perfect'' considering the measurement uncertainties, then we would expect the distribution of $\Delta P_{rot, gyro}$ to be Gaussian, with a width defined by propagating the typical rotation period uncertainties. Thus, in Figure \ref{fig:hist_x} we also show a half-normal distribution (to account for the absolute value factor in the x-parameter) as a black line, and include the corresponding fractions at a given $\sigma$ in Table \ref{tab:results}.

Both in Figure \ref{fig:hist_x} and Table \ref{tab:results}, the Clusters sample exhibits cumulative distributions below the half-normal distribution (e.g., the probability of $x\leq1$ in Table \ref{tab:results} for the half-normal distribution is $0.683$, slightly higher than the values for the Clusters). This shows that, while the Clusters present the best agreement with gyrochronology compared to the other samples, it is below of what we could expect under the ``best-case scenario'' of the Gaussian distribution.

We now examine our results in further detail by comparing the clusters among each other, with respect to the \citet{2019AJ....158..173A} relation. There is a clear trend in Table \ref{tab:results} for the younger Praesepe cluster to show the best agreement with the relation among the clusters, then the older NGC6811, and finally the oldest $\sim 2.5$ Gyr clusters Ruprecht 147 and NGC6819, with the worst agreement. This suggests a possible degradation of the gyrochronology relations beyond the $\sim 1$ Gyr of age, and we further discuss this in Section \ref{sec:degradation}.

\subsubsection{On the inclusion of rotational outliers} \label{sec:discussion_clusters_outliers}

All in all, while the results for clusters are good, they are not perfect. This may be the case because the current gyrochronology relations might not be working as well as hoped, which is the question that motivates the present work. 

But besides that possibility, the partial disagreement we are finding between our sets of coeval stars and the present gyrochronology relations also exists, in part, because not every \textit{bona-fide} member of the clusters should be expected to fall on its corresponding gyrochrone. This may be caused by a number of reasons: rotational outliers, close binaries, stars close to the fully convective regime, or the K-dwarf stalling identified in $\gtrsim 1$ Gyr clusters \citep[e.g.,][]{2019ApJ...879...49C}. Lastly, incorrect rotation period measurements (i.e., measuring a harmonic of the true value) is always a possibility.

In Figure \ref{fig:clusters} there are many cluster members that tend to stray away from the expected gyrochrone (solid lines). These are rotational outliers, with rotation periods mostly shorter than expected. Of these outliers, presumably many are close binaries that survived our HRD cuts, where tidal effects could cause unusual rotation rates with respect to expectations for isolated stars (e.g., \citealt{2022MNRAS.512.2051D, 2022ApJ...930....7A}; see also \citealt{2017ApJ...842...83D} and \citealt{2022MNRAS.513.4380I}). These are known to be relatively numerous in clusters, as the literature reports binary fractions in clusters being around $10\%$ to $40\%$ \citep[e.g.][]{2010MNRAS.401..577S, 2020AJ....159...11C, 2023arXiv230111061D}. In the \citet{2019AJ....158..173A} relation that we use for our test, the outliers are absent, as the authors performed a process of sigma clipping to eliminate them from their calibration on the Praesepe cluster.

For our test, we chose to keep these rotational outliers in the sample because, as Figure \ref{fig:clusters} shows, they tend to be relatively common in clusters, meaning they are likely also common in the field. In this latter case, however, it is very difficult to determine if a given field star is a rotational outlier, as there is no underlying rotational sequence that can be determined for its age, when we, naturally, do not know its age. We do, however, further examine the impact that Gaia-identified close-binaries have in our results (for all the samples) in Section \ref{sec:results_close_binaries}.

Alternatively, in the color regime $G_{BP}-G_{RP} \gtrsim 2$, many of the outliers could be caused by stars near the fully convective regime mentioned in Section \ref{sec:data}. Finally, the stalled spin-down of NGC6811 and the $2.5$ Gyr clusters seen in Figure \ref{fig:clusters} (and described in Section \ref{sec:data_clusters}) is reflected in the worse results shown by these systems in Table \ref{tab:results}, in comparison with the younger Praesepe. \citet{2020A&A...636A..76S} proposed a hypothesis that would explain this effect as a competition between the effects of wind braking and interior coupling in the star (see also \citealt{2023arXiv230107716C}). Seemingly, this so-called ``K-dwarf'' stalling constitutes one of the major challenges for gyrochronology.

\begin{table*}
\centering
\begin{tabular}{|l|@{\hskip 0.25in}rccc@{\hskip 0.25in}|rccc|}
\hline
 &\multicolumn{4}{|c|}{Angus et al. (2019)}  & \multicolumn{4}{c|}{Spada \& Lanzafame (2020)} \\
Data samples & N of Pairs Tested    & $x\leq1$   & $x\leq2$  & $x\leq3$ & N of Pairs Tested  & $x\leq1$  & $x\leq2$  & $x\leq3$  \\ \hline

\textbf{Half-normal distribution} & - & \textbf{0.683} & \textbf{0.955} & \textbf{0.997}  & - & \textbf{0.683} & \textbf{0.955} & \textbf{0.997} \\ \hline

Praesepe & 125x20=2500 & 0.652 & 0.814 & 0.868 & 105x20=2100 & 0.539 & 0.805 & 0.912 \\
NGC6811 & 75x20=1500 & 0.317 & 0.564 & 0.773 & 65x20=1300 & 0.577 & 0.805 & 0.899 \\
Ruprecht 147 + NGC6819 & 31x20=620 & 0.248 & 0.384 & 0.518 & 27x20=540 & 0.370 & 0.602 & 0.746 \\
\textbf{Total Clusters} & \textbf{4620} & \textbf{0.489} & \textbf{0.675} & \textbf{0.790} & \textbf{3940} & \textbf{0.528} & \textbf{0.777} & \textbf{0.885} \\
\hline
\cite{2016MNRAS.455.4212D} & 5 & 0.600 & 0.800 & 0.800 & 3 & 0.667 & 0.667 & 0.667 \\
\cite{2017ApJ...835...75J} & 26 & 0.385 & 0.577 & 0.731 & 11 & 0.182 & 0.273 & 0.727 \\
\cite{2018MNRAS.479.4440G} & 4 & 0.250 & 0.250 & 0.250 & 3 & 0.000 & 0.333 & 0.333 \\
\cite{2018MNRAS.480.4884E} & 5 & 0.200 & 0.600 & 0.800 & 2 & 0.500 & 0.500 & 1.000 \\
\textbf{Total Wide Binaries} & \textbf{40} & \textbf{0.375} & \textbf{0.575} & \textbf{0.700} & \textbf{19} & \textbf{0.263} & \textbf{0.368} & \textbf{0.684} \\
\hline
\textbf{Random Pairs [1]} & \textbf{40x20=800} & \textbf{0.186} & \textbf{0.361} & \textbf{0.500} &  \textbf{19x20=380} & \textbf{0.182} & \textbf{0.366} & \textbf{0.542} \\
\textbf{Random Pairs [2]} & \textbf{11011} & \textbf{0.095} & \textbf{0.182} & \textbf{0.266} & \textbf{8957} & \textbf{0.059} & \textbf{0.113} & \textbf{0.162} \\

\end{tabular}
\caption{Summary of results for the x-parameter for each sample (Equation \ref{eq:x}). The row with \textbf{Total Clusters} presents the whole clusters sample, i.e., the sum of the three rows listed above. The row with \textbf{Total Wide Binaries} presents the Wide Binaries sample, comprised of the four sources listed above. For reference, we also present a row for the results of a standard \textbf{half-normal distribution}. The N of Pairs Tested columns show the size of each sample. For the samples that have been re-sampled, the multiplying factor (x20) is shown for clarity (see Section \ref{sec:data} for details). The $x\leq1,2,3$ columns show the fraction of pairs of each sample that satisfy said equation. Columns 2 through 5 show the results for \citet{2019AJ....158..173A}, and columns 6 to 9 show the results for \citet{2020A&A...636A..76S}. As with Figures \ref{fig:prot_check_hist} and \ref{fig:hist_x}, the Clusters and Wide Binaries tend to show a better agreement with gyrochronology than both random pairs samples. While the two relations yielded fairly similar results, the \citet{2020A&A...636A..76S} relation presents systematically better agreement with gyrochronology across practically all samples. As later discussed in Section \ref{sec:degradation}, this may be caused by the significant differences in the selection filters for the pairs colors and rotation periods. The half-normal distribution and the Random Pairs results represent the upper and lower boundaries, respectively, for the agreement between the data and gyrochronology.}
\label{tab:results}
\end{table*}

\subsection{Random Pairs} \label{sec:discussion_random_pairs}

Both Random Pair [1] and [2] samples show very similar $\Delta P_{rot,gyro}$ distributions in Figure \ref{fig:prot_check_hist}, although the latter one shows a slightly higher peak. Their overall agreement validates both methods of random pair generation. 

However, the differences arise when we compare them in the distributions of the x-parameter in Table \ref{tab:results} and Figure \ref{fig:hist_x}. Although the CDF of $x$ for the Random Pairs [1] raises more rapidly with increasing $x$ than the one for Random Pairs [2], this does not necessarily mean a worse agreement with the gyrochronology relations for the latter sample. This is because the rotation period uncertainties $\sigma_{ P_{rot}}$ of \citet{2014ApJS..211...24M} (the source of the Random Pairs [2] sample) tend to be much lower than the ones from the Wide Binaries sample, and thus for the Random Pairs [1] sample. The median rotation period fractional uncertainty for the Random Pairs [2] is $\sigma_{P_{rot}}/P_{rot} \sim 0.008$, while the median for the Random Pairs [1] sample is $\sim 0.076$, almost and order of magnitude larger. Measurement errors of $\sigma_{P_{rot}}/P_{rot}<0.05$ are likely underestimated and do not account for the uncertainties inherited from phenomena such as differential rotation, spot evolution, and the lifetime of active regions \citep[e.g.,][]{2017MNRAS.472.1618G, 2021MNRAS.508..267S, 2022ApJ...924...31B}. All of these result in much higher values of $x$ for Random Pairs [2] (as the denominator in Equation \ref{eq:x} is smaller), making the comparison of the x-parameter between the Wide Binaries and Random Pairs [1] samples (which have identical $\sigma_{P_{rot}}$ distributions) much more appropriate than with Random Pairs [2]. 

All things considered, both Random Pairs samples show worse agreement with gyrochronology than the Clusters and Wide Binaries. This implies that the standard gyrochronology relations do have some predictive power in identifying coeval populations of field stars.

Ultimately, the results from the Random Pairs samples show a baseline for the non-zero agreement with gyrochronology that can be expected for unassociated pairs of field stars. This could be considered as analogous to the existence of chemical doppelgangers found in studies of Galactic archaeology \citep[e.g.,][]{2018ApJ...853..198N}, although in our case the dimensionality of the agreement is significantly smaller. 

\subsection{Wide Binaries} \label{sec:discussion_binaries}

The Wide Binaries sample shows a better agreement with the tested relations in Figure \ref{fig:prot_check_hist}, in comparison with the samples of random pairs. The distribution of $\Delta P_{rot,gyro}$ is both narrower and taller than the ones from the Random Pairs.  However, it also shows a worse agreement with gyrochronology than the combined Clusters sample.

As already mentioned in Section \ref{sec:discussion_random_pairs}, the best comparison between wide binaries and non-coeval pairs would be that made against the sample of Random Pairs [1], as their $\sigma_{P_{rot}}$ distributions are, by construction, identical. Here, it is shown that the cumulative distribution of $x$ raises substantially more rapidly for the Wide Binaries than for the Random Pairs [1]. Moreover, in Table \ref{tab:results}, we can compare quantitatively the $x$ cumulative distributions for $x=1,2,3$. For instance, for the \citet{2019AJ....158..173A} relation, we have that the ratio of Wide Binaries to Random Pairs [1] that satisfy $x\leq1$ is $\approx 2.0$ ($0.375/0.186$). At face value, our results support empirically the initial assumption that wide binary components are, indeed, coeval.

Although the Wide Binaries sample shows better results than the Random Pairs, its agreement with gyrochronology is clearly below the Clusters sample. As we would expect many of our field binaries to be of several Gyr of age, the effect of weakened magnetic braking in older stars \citep[][]{2016Natur.529..181V, 2021NatAs...5..707H, 2022MNRAS.510.5623M} is likely to be an important factor that affects our sample. By the same line of reasoning, our method should be sensitive to such departures from standard gyrochronology, and therefore it can be used to examine future calibrations that will account for this weakened magnetic braking. Today, given the lack of suitable clusters, only wide binaries \citep[][]{2012ApJ...746..102C}, stars suitable for seismic analyses \citep[][]{2014A&A...572A..34G,2021NatAs...5..707H}, and possibly other seemingly coeval structures \citep[e.g.,][]{2022AJ....163..275A, 2022AJ....164..137K} offer hope to better calibrate gyrochronology at several Gyr. 

\subsection{On the color dependence} \label{sec:refinements}

Gyrochronology relations usually depend directly on color (used as a proxy for stellar mass), such that for a given age, the $P_{rot}$ will increase toward redder colors (lower masses). Given this intrinsic dependence, for a given set of gyrochrones that span a range of ages, the gyrochrones will tend to be more clustered at bluer colors (higher masses), and they will diverge and be more separated at redder colors (e.g., see Figure \ref{fig:clusters}). Thus, bluer stars with very different ages may appear with similar rotation periods. This could introduce a bias in our test, as small uncertainties in the rotation period of a blue primary star can imply a large $\Delta P_{rot,gyro}$ for a redder secondary. 

To examine the effect of this in our tests, we divided each of our samples into two sub-samples of equal size, by small and large color differences between the primary and secondary. We found that, in general, pairs with similar colors showed smaller $\Delta P_{rot,gyro}$ and x-parameter values than pairs with different colors. This happens with both coeval and non-coeval pairs.

To weigh the dependence of our test on the color of the primary, we flagged pairs with primary colors that were bluer than a certain threshold (we tested several limits between $(G_{BP}-G_{RP})_0=0.7$ and 1.0 mag, and the results were very similar), and defined the reference gyrochrone for $\Delta P_{rot,gyro}$ by fitting it to the redder secondary instead. This resulted in $\Delta P_{rot,gyro}$ values that tended to be closer to zero than the original sample, for samples with both coeval and non-coeval star pairs. However, the x-parameter values were almost identical. This is because, by definition (see Section \ref{sec:rot_test}), the x-parameter value is basically independent of which star of the pair is treated as the primary. Thus, at least partially, our x-parameter does take this dependence on the primary's color into account.

Ultimately, our samples are too small to refine our test as to fully take color into account. In order to mitigate the apparent bias in pairs of stars with similar colors, future efforts with larger samples will be needed.

\subsection{The effect of unresolved close-binaries} \label{sec:results_close_binaries}

Following our discussion in Section \ref{sec:discussion_clusters_outliers}, we now assess the impact that close binaries may have in our results. This arises from the possibility that the rotation periods of the stars we study, and therefore their $\Delta P_{rot,gyro}$ and x-parameter values, may be affected by the presence of unresolved companions at close separations that have survived our HRD cuts. We do this by taking advantage of the recently published Gaia EDR3 and DR3 catalogs \citep{2021A&A...649A...1G,2022arXiv220800211G}.

First, we studied the subset of stars with high Renormalized Unit-Weight Error (RUWE), which can be used to identify unresolved companions due to the astrometric noise they imprint in the Gaia solutions \citep{2020MNRAS.496.1922B}. We searched the Gaia DR3 RUWE information for our stars, and found values for virtually 100\% of them. We classified those with RUWE$>1.4$ as having high-RUWE (e.g., see \citealt{2021MNRAS.506.2269E}, \citealt{2020AJ....159..280B}). Across all samples, $\sim$ 20\% of the pairs were classified as having at least one component with high-RUWE. When comparing the pairs' x-parameter distributions, however, we noticed that the ones with high-RUWE components have nearly identical distributions as the full samples. Therefore, we conclude that while these high-RUWE stars may host close companions, these are not having an impact on the rotation periods of the stars we study. This is in agreement with the recent results by \citet{2022ApJ...930....7A} (e.g., see their figure 13).

Second, we studied the subset of stars that can be classified as radial velocity (RV) variables. We again searched in Gaia DR3 and found that $\sim$ 45\% of the pairs in our samples have at least one component with RV data. To identify the RV-variable stars, we followed the prescription of \citet{2022arXiv220605902K} (see their section 3.7). Out of the stars with RV data, we found the fraction of pairs with at least one RV-variable star to be $\sim$ 5\textendash15\% across all samples. When comparing the pairs' x-parameter distributions with their respective full samples, these pairs with RV-variable stars showed noticeably worse agreements with the gyrochronology models. Further inspections showed that these RV-variable stars tend to be concentrated at bluer colors (presumably inherited from the Gaia RV selection function), and more importantly, at shorter rotation periods. For instance, for the most numerous Random Pairs [2] sample, the RV-variable stars predominantly have periods shorter than 5 days. This is again in agreement with \citet{2022ApJ...930....7A}, where spectroscopic binaries are heavily concentrated at short periods. These stars are likely in the regime where tidal interactions are affecting their rotation rates, hence directly affecting the parameters we study (e.g., see also \citealt{2019ApJ...871..174S,2020ApJ...898...76S,2020AJ....160...90A}).

For simplicity, in the analysis that follows we kept all the targets in our full samples. This is motivated by the unnoticeable effect that the high-RUWE stars have in our results, and the small fraction that the RV-variable stars represent compared to the full samples ($\sim$ 3\%). In other words, our conclusions are not impacted by this choice. Nonetheless, we highlight that when applying gyrochronology to field MS stars, or performing analyses similar to ours in larger samples, the existing RV-variability data should be checked as to flag potential rotational outliers (see also \citealt{2022MNRAS.512.2051D}).

\subsection{Comparing the age (dis)agreement directly} \label{sec:results_age}

A final, alternative way of quantifying the agreement of our pairs with gyrochronology, is to directly compare the gyrochronology-derived ages of their members, computed from the relations tested. We followed the method of \citet{2022ApJ...930...36O}, and defined an age ratio ``tolerance'' $\Delta$, so we can test if the ages of the components of a given pair are consistent within a certain $\Delta$. For simplicity, instead of defining distributions for the age ratio given by the gyrochronology equations, we compared directly the computed ages. In summary, we calculated the fraction of pairs in a certain sample that satisfied

\begin{equation}
    1-\Delta < \frac{\text{Age}_{\text{primary}}}{\text{Age}_{\text{secondary}}} < 1+\Delta
\end{equation}

For instance, when adopting a value of $\Delta = 30\%$, we found the following fractions of the samples to be ``coeval'': $\sim 73\%$ for Praesepe, $\sim 46\%$ for NGC6811, $\sim 36\%$ for our 2.5 Gyr clusters, $\sim 50\%$ for our Wide Binaries, $\sim 23\%$ for our Random Pairs [1], and $\sim 25\%$ for our Random Pairs [2]. For a tolerance of $\Delta = 10\%$, we found ``coeval'' fractions of: $\sim 38\%$ for Praesepe, $\sim 20\%$ for NGC6811, $\sim 17\%$ for our 2.5 Gyr clusters, $\sim 20\%$ for our Wide Binaries, $\sim 8\%$ for our Random Pairs [1], and $\sim 8\%$ for our Random Pairs [2]. Note that even in the case of the youngest Praesepe cluster, the agreement is not perfect, as discussed in \ref{sec:discussion_clusters_outliers}.

These results follow a similar trend to the ones shown by the other results in this section, validating our $\Delta P_{rot,gyro}$ and x-parameter tests, and further suggesting a decline in the reliability of gyrochronology for old stars (Section \ref{sec:degradation}).

\begin{figure*}
    \centering
    \includegraphics[width=\linewidth]{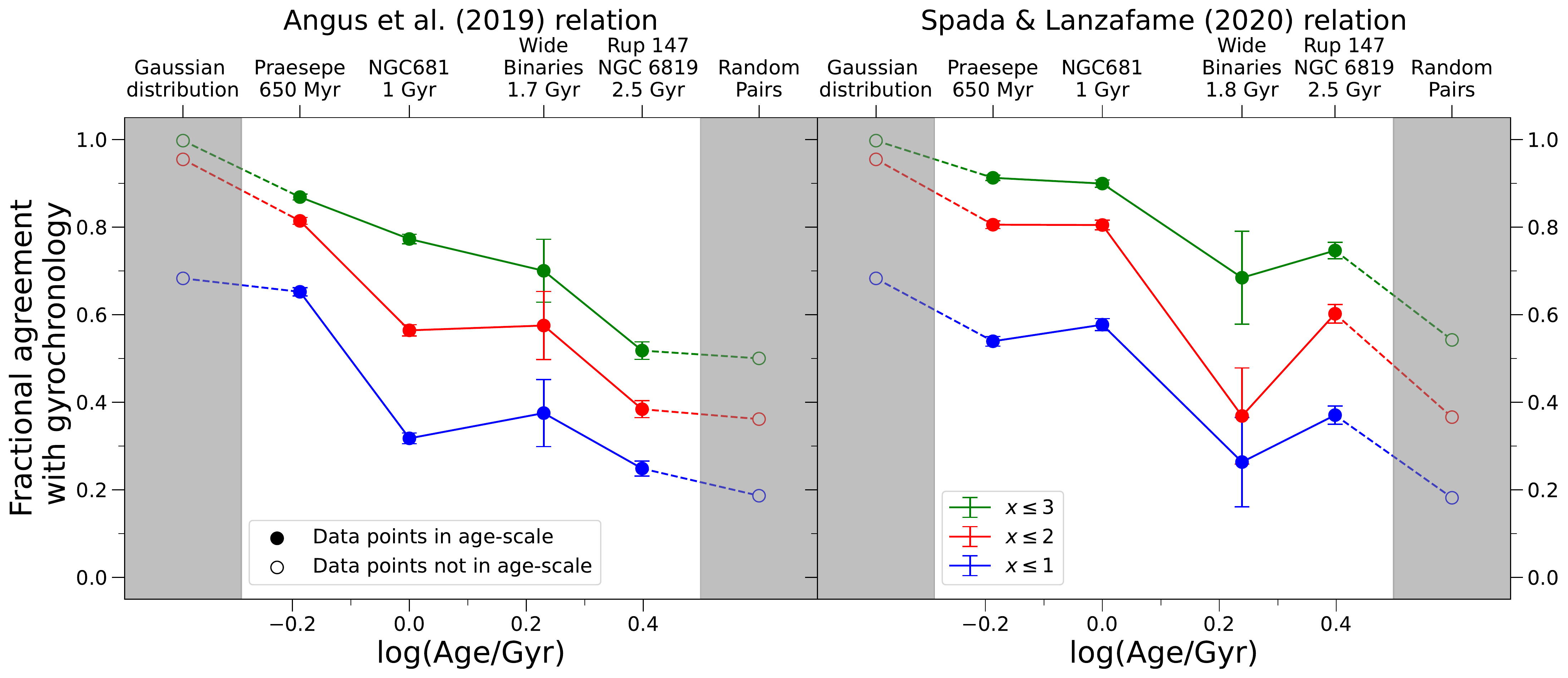}
    \caption{Illustration of how gyrochronology progressively degrades towards older ages: the fraction of pairs that satisfy $x\leq1,2,3$ (in blue, red, green, respectively), as a function of age in logarithmic scale, for the \citet{2019AJ....158..173A} relation (left) and the \citet{2020A&A...636A..76S} relation (right). The error bars correspond to the standard errors calculated using a bootstrap method (see Section \ref{sec:degradation}). The points correspond to (from left to right) the Gaussian distribution, the Praesepe cluster sample, the NGC6811 cluster sample, the Wide Binaries sample, the Ruprecht 147 and NGC6819 combined clusters sample, and the Random Pairs [1] sample. The Gaussian and Random Pairs are plotted for comparison purposes, however, they do not represent real pairs and are not on the same age-scale as the other samples. Thus, we show them with open circles connected by dashed lines, and with a grey background. The breakdown of gyrochronology is illustrated by the older samples showing progressively worse results than the younger ones.}
    \label{fig:age_progression}
\end{figure*}

\begin{figure}
    \centering
    \includegraphics[width=\columnwidth]{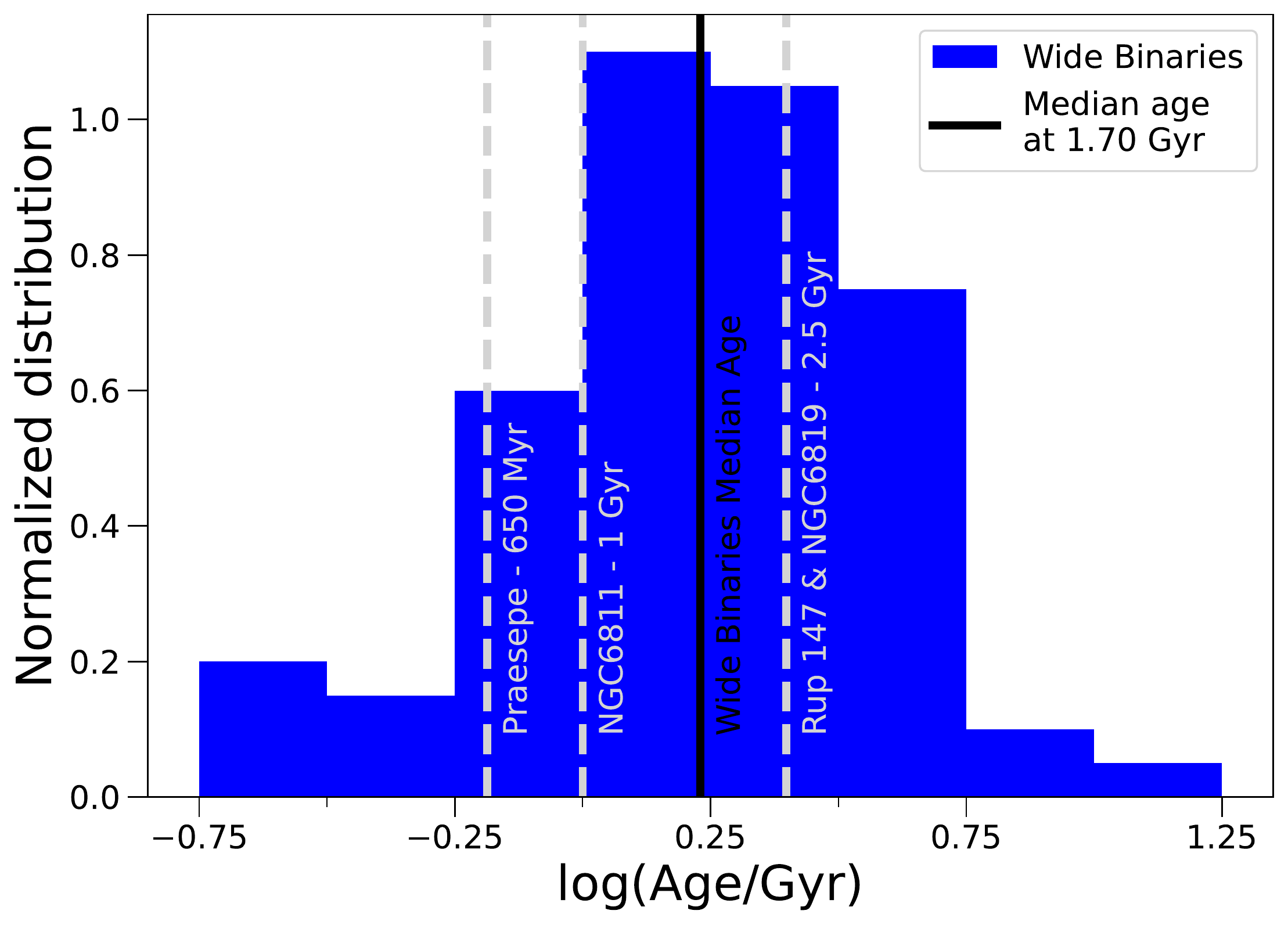}
    \caption{Distribution of the gyrochronology ages of the Wide Binary components (blue), calculated with the \citet{2019AJ....158..173A} relation, in a logarithmic scale. The black vertical line represents the median $\log(\text{age}/\text{Gyr}) = 0.23$, corresponding to $1.7$ Gyr. For comparison purposes, the grey vertical lines show the ages of the open clusters. Note the large age spread for the Wide Binaries.}
    \label{fig:logage}
\end{figure}

\section{The degradation of gyrochronology towards older ages} \label{sec:degradation}

There is mounting evidence that gyrochronology, at least as described with the best current relations, does not fully encapsulate the complex behaviour exhibited by old stars \citep[][]{2016Natur.529..181V, 2019ApJ...871...39M, 2019ApJ...872..128V, 2022ApJ...933..114D}. In order to find out what our data and methodology tell us about this question, in Figure \ref{fig:age_progression} we quantify the fractional agreement between the coeval nature of our cluster and wide binary pairs and the predictions of gyrochronology. We do this for both the \citet{2019AJ....158..173A} and \citet{2020A&A...636A..76S} relations (left- and right-hand panels, respectively), and plot the fractional agreement of each sample at their corresponding age. In Figure \ref{fig:age_progression} we include vertical error bars calculated through a bootstrap method (they are plotted for every point, though may be smaller than the symbols). Note that the large error bar present for the Wide Binaries samples, compared to the clusters', is mainly due to the large difference in sample sizes, as the Wide Binaries sample has fewer pairs than the over-sampled Clusters. The clusters have known ages from their HRDs. To age-rank the Wide Binaries, we use the gyrochronology ages derived in Section \ref{sec:results_age}. These are shown in Figure \ref{fig:logage}, yielding a nearly symmetrical logarithmic distribution with peak and dispersion given by $\log(\text{age}/\text{Gyr}) = 0.23 \pm 0.36$.  Therefore, while the bulk of the sample have gyrochronology ages of $1.5-2.0$ Gyr, there are pairs with gyrochronology ages as young as $200$ Myr and as old as $5$ Gyr.  Finally, in both panels of Figure \ref{fig:age_progression} we include the gyrochronology agreement of the half-normal and Random Pairs [1] distributions (as done in Figure \ref{fig:hist_x}).  As discussed in Section \ref{sec:discussion}, these represent the expected upper and lower boundaries of the agreement with gyrochronology within our samples, respectively. Thus, the Gaussian and Random Pairs [1] samples, while not on the age scale, represent the ``best-'' and ``worst-case scenario'' for the agreement with gyrochronology.

The two panels of Figure \ref{fig:age_progression} show a clear degradation of this ``agreement with gyrochronology'' as we go from younger to older samples.  This is a solid result of our $\Delta P_{rot,gyro}$ statistic, independent of which gyrochronology relation is examined, and further supports existing literature that suggests a significant deviation from a standard Skumanich-like law. 

In order to further explore the trend seen in Figure \ref{fig:age_progression}, we tried to filter our Wide Binaries sample by age, using three different works in the literature:

\begin{itemize}

    \item \citet{2020MNRAS.496.5176D} performed a search of wide binary candidates on the clusters Alpha Per ($85$ Myr old), Pleiades ($125$ Myr), and Praesepe ($650$ Myr). As these stars sample young ages compared to our wide binary sample, we aimed at comparing their gyrochronology agreement versus the one for our sample of field binaries. We searched for rotation periods for their young binary candidates in the literature, but this yielded an extremely small number of pairs, preventing us from extending the analysis to ages younger than Praesepe.
    
    \item  \citet{2021AJ....161..189L} grouped \textit{Kepler}-field stars by fundamental properties ($M_G$, $T_{eff}$, $P_{rot}$ and Rossby number, the ratio of rotation period to the convective overturn timescale), and then applied an age-velocity-dispersion relation to estimate kinematic ages for tens of thousands of stars with measured rotation periods. We performed a cross-match between these stars and our Wide Binaries sample, to again compile a sub-sample of pairs with ages determined independently of gyrochronology. Our matches, however, were too few to allow any significant inferences.
    
    \item \citet{2021AJ....162..100C} constructed a catalog of over 30,000 \textit{Kepler} stars with kinematic properties, grouping them as part of the thin disk (younger), thick disk (older), Hercules stream and halo. As thin disk stars are younger on average than thick disk stars, we tried a cross-match with our Wide Binaries, but found only thin disk stars. 
    
    Therefore, to further probe the age dependence of gyrochronology as depicted in the left panel of Figure \ref{fig:age_progression} with only wide binaries, larger samples of these systems that can be separated in age ranges, ideally with methods independent of gyrochronology, are needed.

 \end{itemize}

The apparent degradation of gyrochronology for older stars served as a motivation for the semi-theoretical model from \citet{2020A&A...636A..76S}. Indeed, at a first look, in the right panel of Figure \ref{fig:age_progression}, the \citet{2020A&A...636A..76S} relation tends to show a  better agreement with gyrochronology than the \citet{2019AJ....158..173A} relation for almost all samples. The improvement is most obvious for the $2.5$ Gyr clusters, which have been notoriously difficult to describe with the usual relations calibrated with younger clusters \citep[e.g.][]{2020ApJ...904..140C}.  The most significant difference between both relations in Figure \ref{fig:age_progression}, is the results of the Wide Binaries in comparison to the other samples. This can be at least partially explained by the rotation period and color constraints we used to construct our samples (see Section \ref{sec:data_clusters}), as they ended up being much more restrictive in the \citet{2020A&A...636A..76S} than for the \citet{2019AJ....158..173A} relation. This was due to the tighter color distribution and smaller age range allowed by the \citet{2020A&A...636A..76S} model. In practice, these meant that we could only test pairs with ages $\leq 4.57$ Gyr (i.e., stars with rotation periods below their oldest gyrochrone), which discarded more than half of the original Wide Binary sample used for \citet{2019AJ....158..173A}. Further, these constraints resulted in vastly different pair distributions in the rotation period versus color diagram. In the end, due to the considerable differences in the construction of the samples, the results for the Wide Binaries with the \citet{2019AJ....158..173A} versus the \citet{2020A&A...636A..76S} relation are not directly comparable. 

In spite of the above, the trend of degradation of the examined gyrochronology models towards older ages is evident with the Clusters samples alone, in the two panels of Figure \ref{fig:age_progression}. Particularly, noticing that the $\sim$ 1 and 2.5 Gyr old clusters show a significantly better agreement for the \cite{2020A&A...636A..76S} model, it can be argued that the results of Figure \ref{fig:age_progression} validate their treatment of the stalled spin-down.

In conclusion, our results point out a clear need for improved  gyrochronology relations beyond $\sim 1$ Gyr. This overall trend is in qualitative agreement with other findings in the literature, particularly that the rotational distributions of \textit{Kepler}-field stars cannot be explained with standard angular momentum evolution (\citealt{2015MNRAS.450.1787A}; \citealt{2016Natur.529..181V}; see also \citealt{2021NatAs...5..707H}; \citealt{2022MNRAS.510.5623M}). This strengthens the need for better calibrations at older ages, and may be hinting towards a more restricted validity range of gyrochronology than has been hoped for, or rather that rotation becomes, inevitably, a less reliable stellar-chronometer as stars age. Ultimately, novel constraints at old ages will be needed to resolve these issues.

\section{Conclusions}\label{sec:conclusions}

Gyrochronology is a promising age-dating technique for field MS stars, but its reliability beyond ages of a few Gyr is uncertain. In this paper, we designed a dedicated quantitative method to test existing gyrochronology models in the regime of ages younger than $\sim 2.5$ Gyr using pairs of coeval stars, in the form of both wide binaries and pairs of cluster members.  The test is based on the assumption of the coeval nature of the components of said systems.  Samples of pairs of stars with coeval components and well-measured rotation periods were assembled and then tested against three gyrochronology relations: \citet{2019AJ....158..173A}, \citet{2015MNRAS.450.1787A}, and \citet{2020A&A...636A..76S}.
Two samples of random pairs were constructed in order to have sets of pairs with non-coeval components that serve as comparison with the coeval pairs.  The test consisted of calculating the difference between the measured rotation period of the secondary of each pair and its expected rotation period given the age of the primary star estimated via gyrochronology (see Figure \ref{fig:rot_eg}). Then, we compared this difference in rotation period with its propagated measurement error, which we call the x-parameter (Equation \ref{eq:x}).

Even though the gyrochronology relations we tested were calibrated with star
cluster samples that filtered out rotational outliers, the samples that we
constructed for this work are designed such that we can test the performance
of the gyrochronology relations when applied to the most general case of
isolated (in the sense that not in a cluster) field stars.  These field
stars, in general, cannot be identified a priori as rotational outliers
based on their rotation periods, as these are (along with their colors)
precisely what will be used to determine the agreement or disagreement
with a gyrochronology relation at a given age (which is the purpose of
our test).  For this reason, unlike the works that calibrated the
relations, we did not exclude from our samples stars that, based on their
location on the $P_{rot}$-color plane, may be considered rotational outliers.

Many of these problematic objects are close binaries and evolved stars,
which become rotational outliers due to the effect of tidal interactions
and stellar expansion, respectively, on the angular momentum evolution.
However, thanks to Gaia's precision parallaxes we can avoid a fraction
(not all) of these potential outliers, independently of their rotational
properties, by appropriate cuts in the HRD. We therefore removed from our
samples all stars in the close binary sequence just above the MS, and did
the same for evolved stars by enforcing an upper limit on absolute
magnitude. We also tested the effects that close, unresolved-companions have in stellar rotation rates by examining the Gaia DR3 RUWE and RV-variability parameters. We found that stars with high RUWE do not have any significant effect on the results of our test.  On the other hand, stars flagged as RV-variable were few in our samples, but close examination revealed that these tend to bias the measured rotation periods towards more rapid rotation rates, thus increasing the difference with the expectations from gyrochronology.  However, these RV-variables are a near negligible fraction of our stars, and removing or keeping them in our samples does not affect the final results.

Overall, the results showed a better agreement between the data and the state-of-the-art gyrochronology relations for the wide binaries than for the random pairs, and an even better agreement for the clusters (Figures \ref{fig:prot_check_hist} and \ref{fig:hist_x}). However, not even the clusters that were used in the calibration of the relations being tested achieve a perfect agreement as quantified by our test. This may be explained by the presence of said rotational outliers.

Further explorations showed that the agreement between the data for coeval stars and the gyrochronology relations gets progressively worse with increasing age (Figure \ref{fig:age_progression}), reflecting a degradation of the predictions of standard gyrochronology models for stars older than $\sim 1$ Gyr. 

Therefore, while the present gyrochronology predictions appear to be in reasonable -- though not extraordinary -- concordance with all the samples of pairs of coeval stars we have studied (i.e., clusters and field wide binaries), our results show that the current form of gyrochronology needs improvements and/or a better understanding of low-mass stellar spindown as a function of time, particularly when reaching the Gyr-timescales.

In summary, we have presented a technique that should allow a useful empirical test of age-rotation relations in the era of ever-increasing space-based photometry.  We highlight that the underlying technique used here is independent of the specific age-rotation relations used. Furthermore, this test can be easily adapted to empirically assess and compare the different gyrochronology relations, and evaluate the effectiveness of new equations and functional forms that may be published in the future. Moreover, it can be adapted to activity-age relations and other age determination techniques for field-stars. This represents another step in the direction of reliable and precise age determination for field MS stars. 

\section*{Acknowledgements}
We thank Jennifer van Saders, \^{A}ngela Santos, Savita Mathur, Erik Mamajek, Jamie Tayar, David Soderblom, and Sean Matt for helpful remarks. We also thank Marc Pinsonneault and the Ohio State stars group for insightful comments and discussions. We also thank Ruth Angus and Alessandro Lanzafame for helpful remarks regarding their respective gyrochronology relations. Finally, we thank the referee, Terry Oswalt, for a thorough review of our manuscript, which led to improvements and clarity.

DGR acknowledges support from the Spanish Ministry of Science and Innovation (MICINN) with the grant no. PID2019-107187GB-I00. JC acknowledges support from the Agencia Nacional de Investigación y Desarrollo (ANID) via Proyecto Fondecyt Regular 1191366; and from ANID BASAL projects CATA-Puente ACE210002 and CATA2-FB210003.

This work has made use of data from the European Space Agency (ESA) mission
{\it Gaia} (\url{https://www.cosmos.esa.int/gaia}), processed by the {\it Gaia}
Data Processing and Analysis Consortium (DPAC,
\url{https://www.cosmos.esa.int/web/gaia/dpac/consortium}). Funding for the DPAC
has been provided by national institutions, in particular the institutions
participating in the {\it Gaia} Multilateral Agreement.

\section*{Data Availability}

The data and code used to produce the results of this article are available at \url{https://github.com/jsilvabeyer/wbgyro23} and the online supplementary material.




\bibliographystyle{mnras}
\bibliography{main} 

\begin{thebibliography}{}
\makeatletter
\relax
\def\mn@urlcharsother{\let\do\@makeother \do\$\do\&\do\#\do\^\do\_\do\%\do\~}
\def\mn@doi{\begingroup\mn@urlcharsother \@ifnextchar [ {\mn@doi@}
  {\mn@doi@[]}}
\def\mn@doi@[#1]#2{\def\@tempa{#1}\ifx\@tempa\@empty \href
  {http://dx.doi.org/#2} {doi:#2}\else \href {http://dx.doi.org/#2} {#1}\fi
  \endgroup}
\def\mn@eprint#1#2{\mn@eprint@#1:#2::\@nil}
\def\mn@eprint@arXiv#1{\href {http://arxiv.org/abs/#1} {{\tt arXiv:#1}}}
\def\mn@eprint@dblp#1{\href {http://dblp.uni-trier.de/rec/bibtex/#1.xml}
  {dblp:#1}}
\def\mn@eprint@#1:#2:#3:#4\@nil{\def\@tempa {#1}\def\@tempb {#2}\def\@tempc
  {#3}\ifx \@tempc \@empty \let \@tempc \@tempb \let \@tempb \@tempa \fi \ifx
  \@tempb \@empty \def\@tempb {arXiv}\fi \@ifundefined
  {mn@eprint@\@tempb}{\@tempb:\@tempc}{\expandafter \expandafter \csname
  mn@eprint@\@tempb\endcsname \expandafter{\@tempc}}}

\bibitem[\protect\citeauthoryear{{Ag{\"u}eros} et~al.,}{{Ag{\"u}eros}
  et~al.}{2018}]{2018ApJ...862...33A}
{Ag{\"u}eros} M.~A.,  et~al., 2018, \mn@doi [\apj] {10.3847/1538-4357/aac6ed},
  \href {https://ui.adsabs.harvard.edu/abs/2018ApJ...862...33A} {862, 33}

\bibitem[\protect\citeauthoryear{{Am{\^o}res} et~al.,}{{Am{\^o}res}
  et~al.}{2021}]{2021MNRAS.508.1788A}
{Am{\^o}res} E.~B.,  et~al., 2021, \mn@doi [\mnras] {10.1093/mnras/stab2248},
  \href {https://ui.adsabs.harvard.edu/abs/2021MNRAS.508.1788A} {508, 1788}

\bibitem[\protect\citeauthoryear{{Andrae} et~al.,}{{Andrae}
  et~al.}{2018}]{2018A&A...616A...8A}
{Andrae} R.,  et~al., 2018, \mn@doi [\aap] {10.1051/0004-6361/201732516}, \href
  {https://ui.adsabs.harvard.edu/abs/2018A&A...616A...8A} {616, A8}

\bibitem[\protect\citeauthoryear{{Andrews}, {Chanam{\'e}}  \&
  {Ag{\"u}eros}}{{Andrews} et~al.}{2017}]{2017MNRAS.472..675A}
{Andrews} J.~J.,  {Chanam{\'e}} J.,   {Ag{\"u}eros} M.~A.,  2017, \mn@doi
  [\mnras] {10.1093/mnras/stx2000}, \href
  {https://ui.adsabs.harvard.edu/abs/2017MNRAS.472..675A} {472, 675}

\bibitem[\protect\citeauthoryear{{Andrews}, {Anguiano}, {Chanam{\'e}},
  {Ag{\"u}eros}, {Lewis}, {Hayes}  \& {Majewski}}{{Andrews}
  et~al.}{2019}]{2019ApJ...871...42A}
{Andrews} J.~J.,  {Anguiano} B.,  {Chanam{\'e}} J.,  {Ag{\"u}eros} M.~A.,
  {Lewis} H.~M.,  {Hayes} C.~R.,   {Majewski} S.~R.,  2019, \mn@doi [\apj]
  {10.3847/1538-4357/aaf502}, \href
  {https://ui.adsabs.harvard.edu/abs/2019ApJ...871...42A} {871, 42}

\bibitem[\protect\citeauthoryear{{Andrews}, {Curtis}, {Chanam{\'e}},
  {Ag{\"u}eros}, {Schuler}, {Kounkel}  \& {Covey}}{{Andrews}
  et~al.}{2022}]{2022AJ....163..275A}
{Andrews} J.~J.,  {Curtis} J.~L.,  {Chanam{\'e}} J.,  {Ag{\"u}eros} M.~A.,
  {Schuler} S.~C.,  {Kounkel} M.,   {Covey} K.~R.,  2022, \mn@doi [\aj]
  {10.3847/1538-3881/ac6952}, \href
  {https://ui.adsabs.harvard.edu/abs/2022AJ....163..275A} {163, 275}

\bibitem[\protect\citeauthoryear{{Angus}, {Aigrain}, {Foreman-Mackey}  \&
  {McQuillan}}{{Angus} et~al.}{2015}]{2015MNRAS.450.1787A}
{Angus} R.,  {Aigrain} S.,  {Foreman-Mackey} D.,   {McQuillan} A.,  2015,
  \mn@doi [\mnras] {10.1093/mnras/stv423}, \href
  {https://ui.adsabs.harvard.edu/abs/2015MNRAS.450.1787A} {450, 1787}

\bibitem[\protect\citeauthoryear{{Angus} et~al.,}{{Angus}
  et~al.}{2019}]{2019AJ....158..173A}
{Angus} R.,  et~al., 2019, \mn@doi [\aj] {10.3847/1538-3881/ab3c53}, \href
  {https://ui.adsabs.harvard.edu/abs/2019AJ....158..173A} {158, 173}

\bibitem[\protect\citeauthoryear{{Angus} et~al.,}{{Angus}
  et~al.}{2020}]{2020AJ....160...90A}
{Angus} R.,  et~al., 2020, \mn@doi [\aj] {10.3847/1538-3881/ab91b2}, \href
  {https://ui.adsabs.harvard.edu/abs/2020AJ....160...90A} {160, 90}

\bibitem[\protect\citeauthoryear{{Avallone} et~al.,}{{Avallone}
  et~al.}{2022}]{2022ApJ...930....7A}
{Avallone} E.~A.,  et~al., 2022, \mn@doi [\apj] {10.3847/1538-4357/ac60a1},
  \href {https://ui.adsabs.harvard.edu/abs/2022ApJ...930....7A} {930, 7}

\bibitem[\protect\citeauthoryear{{Bailer-Jones}, {Rybizki}, {Fouesneau},
  {Mantelet}  \& {Andrae}}{{Bailer-Jones} et~al.}{2018}]{2018AJ....156...58B}
{Bailer-Jones} C.~A.~L.,  {Rybizki} J.,  {Fouesneau} M.,  {Mantelet} G.,
  {Andrae} R.,  2018, \mn@doi [\aj] {10.3847/1538-3881/aacb21}, \href
  {https://ui.adsabs.harvard.edu/abs/2018AJ....156...58B} {156, 58}

\bibitem[\protect\citeauthoryear{{Barnes}}{{Barnes}}{2003}]{2003ApJ...586..464B}
{Barnes} S.~A.,  2003, \mn@doi [\apj] {10.1086/367639}, \href
  {https://ui.adsabs.harvard.edu/abs/2003ApJ...586..464B} {586, 464}

\bibitem[\protect\citeauthoryear{{Barnes}}{{Barnes}}{2007}]{2007ApJ...669.1167B}
{Barnes} S.~A.,  2007, \mn@doi [\apj] {10.1086/519295}, \href
  {https://ui.adsabs.harvard.edu/abs/2007ApJ...669.1167B} {669, 1167}

\bibitem[\protect\citeauthoryear{{Barrientos} \& {Chanam{\'e}}}{{Barrientos} \&
  {Chanam{\'e}}}{2021}]{2021ApJ...923..181B}
{Barrientos} M.,  {Chanam{\'e}} J.,  2021, \mn@doi [\apj]
  {10.3847/1538-4357/ac2f49}, \href
  {https://ui.adsabs.harvard.edu/abs/2021ApJ...923..181B} {923, 181}

\bibitem[\protect\citeauthoryear{{Basri}, {Streichenberger}, {McWard},
  {Edmond}, {Tan}, {Lee}  \& {Melton}}{{Basri}
  et~al.}{2022}]{2022ApJ...924...31B}
{Basri} G.,  {Streichenberger} T.,  {McWard} C.,  {Edmond} Lawrence I.,  {Tan}
  J.,  {Lee} M.,   {Melton} T.,  2022, \mn@doi [\apj]
  {10.3847/1538-4357/ac3420}, \href
  {https://ui.adsabs.harvard.edu/abs/2022ApJ...924...31B} {924, 31}

\bibitem[\protect\citeauthoryear{{Belokurov} et~al.,}{{Belokurov}
  et~al.}{2020}]{2020MNRAS.496.1922B}
{Belokurov} V.,  et~al., 2020, \mn@doi [\mnras] {10.1093/mnras/staa1522}, \href
  {https://ui.adsabs.harvard.edu/abs/2020MNRAS.496.1922B} {496, 1922}

\bibitem[\protect\citeauthoryear{{Berger}, {Huber}, {van Saders}, {Gaidos},
  {Tayar}  \& {Kraus}}{{Berger} et~al.}{2020a}]{2020AJ....159..280B}
{Berger} T.~A.,  {Huber} D.,  {van Saders} J.~L.,  {Gaidos} E.,  {Tayar} J.,
  {Kraus} A.~L.,  2020a, \mn@doi [\aj] {10.3847/1538-3881/159/6/280}, \href
  {https://ui.adsabs.harvard.edu/abs/2020AJ....159..280B} {159, 280}

\bibitem[\protect\citeauthoryear{{Berger}, {Huber}, {Gaidos}, {van Saders}  \&
  {Weiss}}{{Berger} et~al.}{2020b}]{2020AJ....160..108B}
{Berger} T.~A.,  {Huber} D.,  {Gaidos} E.,  {van Saders} J.~L.,   {Weiss}
  L.~M.,  2020b, \mn@doi [\aj] {10.3847/1538-3881/aba18a}, \href
  {https://ui.adsabs.harvard.edu/abs/2020AJ....160..108B} {160, 108}

\bibitem[\protect\citeauthoryear{{Bonfanti}, {Ortolani}  \&
  {Nascimbeni}}{{Bonfanti} et~al.}{2016}]{2016A&A...585A...5B}
{Bonfanti} A.,  {Ortolani} S.,   {Nascimbeni} V.,  2016, \mn@doi [\aap]
  {10.1051/0004-6361/201527297}, \href
  {https://ui.adsabs.harvard.edu/abs/2016A&A...585A...5B} {585, A5}

\bibitem[\protect\citeauthoryear{{Borucki} et~al.,}{{Borucki}
  et~al.}{2010}]{2010Sci...327..977B}
{Borucki} W.~J.,  et~al., 2010, \mn@doi [Science] {10.1126/science.1185402},
  \href {https://ui.adsabs.harvard.edu/abs/2010Sci...327..977B} {327, 977}

\bibitem[\protect\citeauthoryear{{Bossini} et~al.,}{{Bossini}
  et~al.}{2019}]{2019A&A...623A.108B}
{Bossini} D.,  et~al., 2019, \mn@doi [\aap] {10.1051/0004-6361/201834693},
  \href {https://ui.adsabs.harvard.edu/abs/2019A&A...623A.108B} {623, A108}

\bibitem[\protect\citeauthoryear{{Bouma}, {Palumbo}  \& {Hillenbrand}}{{Bouma}
  et~al.}{2023}]{2023ApJ...947L...3B}
{Bouma} L.~G.,  {Palumbo} E.~K.,   {Hillenbrand} L.~A.,  2023, \mn@doi [\apjl]
  {10.3847/2041-8213/acc589}, \href
  {https://ui.adsabs.harvard.edu/abs/2023ApJ...947L...3B} {947, L3}

\bibitem[\protect\citeauthoryear{{Bressan}, {Marigo}, {Girardi}, {Salasnich},
  {Dal Cero}, {Rubele}  \& {Nanni}}{{Bressan}
  et~al.}{2012}]{2012MNRAS.427..127B}
{Bressan} A.,  {Marigo} P.,  {Girardi} L.,  {Salasnich} B.,  {Dal Cero} C.,
  {Rubele} S.,   {Nanni} A.,  2012, \mn@doi [\mnras]
  {10.1111/j.1365-2966.2012.21948.x}, \href
  {https://ui.adsabs.harvard.edu/abs/2012MNRAS.427..127B} {427, 127}

\bibitem[\protect\citeauthoryear{{Brice{\~n}o-Morales} \&
  {Chanam{\'e}}}{{Brice{\~n}o-Morales} \&
  {Chanam{\'e}}}{2023}]{2023MNRAS.522.1288B}
{Brice{\~n}o-Morales} G.,  {Chanam{\'e}} J.,  2023, \mn@doi [\mnras]
  {10.1093/mnras/stad608}, \href
  {https://ui.adsabs.harvard.edu/abs/2023MNRAS.522.1288B} {522, 1288}

\bibitem[\protect\citeauthoryear{{Cao}, {Pinsonneault}  \& {van Saders}}{{Cao}
  et~al.}{2023}]{2023arXiv230107716C}
{Cao} L.,  {Pinsonneault} M.~H.,   {van Saders} J.~L.,  2023, \mn@doi [arXiv
  e-prints] {10.48550/arXiv.2301.07716}, \href
  {https://ui.adsabs.harvard.edu/abs/2023arXiv230107716C} {p. arXiv:2301.07716}

\bibitem[\protect\citeauthoryear{{Chanam{\'e}} \& {Gould}}{{Chanam{\'e}} \&
  {Gould}}{2004}]{2004ApJ...601..289C}
{Chanam{\'e}} J.,  {Gould} A.,  2004, \mn@doi [\apj] {10.1086/380442}, \href
  {https://ui.adsabs.harvard.edu/abs/2004ApJ...601..289C} {601, 289}

\bibitem[\protect\citeauthoryear{{Chanam{\'e}} \& {Ram{\'\i}rez}}{{Chanam{\'e}}
  \& {Ram{\'\i}rez}}{2012}]{2012ApJ...746..102C}
{Chanam{\'e}} J.,  {Ram{\'\i}rez} I.,  2012, \mn@doi [\apj]
  {10.1088/0004-637X/746/1/102}, \href
  {https://ui.adsabs.harvard.edu/abs/2012ApJ...746..102C} {746, 102}

\bibitem[\protect\citeauthoryear{{Chaplin} et~al.,}{{Chaplin}
  et~al.}{2014}]{2014ApJS..210....1C}
{Chaplin} W.~J.,  et~al., 2014, \mn@doi [\apjs] {10.1088/0067-0049/210/1/1},
  \href {https://ui.adsabs.harvard.edu/abs/2014ApJS..210....1C} {210, 1}

\bibitem[\protect\citeauthoryear{{Chen} et~al.,}{{Chen}
  et~al.}{2021}]{2021AJ....162..100C}
{Chen} D.-C.,  et~al., 2021, \mn@doi [\aj] {10.3847/1538-3881/ac0f08}, \href
  {https://ui.adsabs.harvard.edu/abs/2021AJ....162..100C} {162, 100}

\bibitem[\protect\citeauthoryear{{Cohen}, {Geller}  \& {von Hippel}}{{Cohen}
  et~al.}{2020}]{2020AJ....159...11C}
{Cohen} R.~E.,  {Geller} A.~M.,   {von Hippel} T.,  2020, \mn@doi [\aj]
  {10.3847/1538-3881/ab59d7}, \href
  {https://ui.adsabs.harvard.edu/abs/2020AJ....159...11C} {159, 11}

\bibitem[\protect\citeauthoryear{{Connelly}, {Bizzarro}, {Krot}, {Nordlund},
  {Wielandt}  \& {Ivanova}}{{Connelly} et~al.}{2012}]{2012Sci...338..651C}
{Connelly} J.~N.,  {Bizzarro} M.,  {Krot} A.~N.,  {Nordlund} {\r{A}}.,
  {Wielandt} D.,   {Ivanova} M.~A.,  2012, \mn@doi [Science]
  {10.1126/science.1226919}, \href
  {https://ui.adsabs.harvard.edu/abs/2012Sci...338..651C} {338, 651}

\bibitem[\protect\citeauthoryear{{Curtis}, {Ag{\"u}eros}, {Mamajek}, {Wright}
  \& {Cummings}}{{Curtis} et~al.}{2019a}]{2019AJ....158...77C}
{Curtis} J.~L.,  {Ag{\"u}eros} M.~A.,  {Mamajek} E.~E.,  {Wright} J.~T.,
  {Cummings} J.~D.,  2019a, \mn@doi [\aj] {10.3847/1538-3881/ab2899}, \href
  {https://ui.adsabs.harvard.edu/abs/2019AJ....158...77C} {158, 77}

\bibitem[\protect\citeauthoryear{{Curtis}, {Ag{\"u}eros}, {Douglas}  \&
  {Meibom}}{{Curtis} et~al.}{2019b}]{2019ApJ...879...49C}
{Curtis} J.~L.,  {Ag{\"u}eros} M.~A.,  {Douglas} S.~T.,   {Meibom} S.,  2019b,
  \mn@doi [\apj] {10.3847/1538-4357/ab2393}, \href
  {https://ui.adsabs.harvard.edu/abs/2019ApJ...879...49C} {879, 49}

\bibitem[\protect\citeauthoryear{{Curtis} et~al.,}{{Curtis}
  et~al.}{2020}]{2020ApJ...904..140C}
{Curtis} J.~L.,  et~al., 2020, \mn@doi [\apj] {10.3847/1538-4357/abbf58}, \href
  {https://ui.adsabs.harvard.edu/abs/2020ApJ...904..140C} {904, 140}

\bibitem[\protect\citeauthoryear{{Daher} et~al.,}{{Daher}
  et~al.}{2022}]{2022MNRAS.512.2051D}
{Daher} C.~M.,  et~al., 2022, \mn@doi [\mnras] {10.1093/mnras/stac590}, \href
  {https://ui.adsabs.harvard.edu/abs/2022MNRAS.512.2051D} {512, 2051}

\bibitem[\protect\citeauthoryear{{David}, {Angus}, {Curtis}, {van Saders},
  {Colman}, {Contardo}, {Lu}  \& {Zinn}}{{David}
  et~al.}{2022}]{2022ApJ...933..114D}
{David} T.~J.,  {Angus} R.,  {Curtis} J.~L.,  {van Saders} J.~L.,  {Colman}
  I.~L.,  {Contardo} G.,  {Lu} Y.,   {Zinn} J.~C.,  2022, \mn@doi [\apj]
  {10.3847/1538-4357/ac6dd3}, \href
  {https://ui.adsabs.harvard.edu/abs/2022ApJ...933..114D} {933, 114}

\bibitem[\protect\citeauthoryear{{Deacon} \& {Kraus}}{{Deacon} \&
  {Kraus}}{2020}]{2020MNRAS.496.5176D}
{Deacon} N.~R.,  {Kraus} A.~L.,  2020, \mn@doi [\mnras]
  {10.1093/mnras/staa1877}, \href
  {https://ui.adsabs.harvard.edu/abs/2020MNRAS.496.5176D} {496, 5176}

\bibitem[\protect\citeauthoryear{{Deacon} et~al.,}{{Deacon}
  et~al.}{2016}]{2016MNRAS.455.4212D}
{Deacon} N.~R.,  et~al., 2016, \mn@doi [\mnras] {10.1093/mnras/stv2132}, \href
  {https://ui.adsabs.harvard.edu/abs/2016MNRAS.455.4212D} {455, 4212}

\bibitem[\protect\citeauthoryear{{Donada} et~al.,}{{Donada}
  et~al.}{2023}]{2023arXiv230111061D}
{Donada} J.,  et~al., 2023, \mn@doi [arXiv e-prints]
  {10.48550/arXiv.2301.11061}, \href
  {https://ui.adsabs.harvard.edu/abs/2023arXiv230111061D} {p. arXiv:2301.11061}

\bibitem[\protect\citeauthoryear{{Douglas}, {Ag{\"u}eros}, {Covey}  \&
  {Kraus}}{{Douglas} et~al.}{2017}]{2017ApJ...842...83D}
{Douglas} S.~T.,  {Ag{\"u}eros} M.~A.,  {Covey} K.~R.,   {Kraus} A.,  2017,
  \mn@doi [\apj] {10.3847/1538-4357/aa6e52}, \href
  {https://ui.adsabs.harvard.edu/abs/2017ApJ...842...83D} {842, 83}

\bibitem[\protect\citeauthoryear{{El-Badry} \& {Rix}}{{El-Badry} \&
  {Rix}}{2018}]{2018MNRAS.480.4884E}
{El-Badry} K.,  {Rix} H.-W.,  2018, \mn@doi [\mnras] {10.1093/mnras/sty2186},
  \href {https://ui.adsabs.harvard.edu/abs/2018MNRAS.480.4884E} {480, 4884}

\bibitem[\protect\citeauthoryear{{El-Badry}, {Rix}  \& {Heintz}}{{El-Badry}
  et~al.}{2021}]{2021MNRAS.506.2269E}
{El-Badry} K.,  {Rix} H.-W.,   {Heintz} T.~M.,  2021, \mn@doi [\mnras]
  {10.1093/mnras/stab323}, \href
  {https://ui.adsabs.harvard.edu/abs/2021MNRAS.506.2269E} {506, 2269}

\bibitem[\protect\citeauthoryear{{Elliott} \& {Bayo}}{{Elliott} \&
  {Bayo}}{2016}]{2016MNRAS.459.4499E}
{Elliott} P.,  {Bayo} A.,  2016, \mn@doi [\mnras] {10.1093/mnras/stw926}, \href
  {https://ui.adsabs.harvard.edu/abs/2016MNRAS.459.4499E} {459, 4499}

\bibitem[\protect\citeauthoryear{{Espinoza-Rojas}, {Chanam{\'e}}, {Jofr{\'e}}
  \& {Casamiquela}}{{Espinoza-Rojas} et~al.}{2021}]{2021ApJ...920...94E}
{Espinoza-Rojas} F.,  {Chanam{\'e}} J.,  {Jofr{\'e}} P.,   {Casamiquela} L.,
  2021, \mn@doi [\apj] {10.3847/1538-4357/ac15fd}, \href
  {https://ui.adsabs.harvard.edu/abs/2021ApJ...920...94E} {920, 94}

\bibitem[\protect\citeauthoryear{{Gaia Collaboration} et~al.,}{{Gaia
  Collaboration} et~al.}{2018}]{2018A&A...616A...1G}
{Gaia Collaboration} et~al., 2018, \mn@doi [\aap]
  {10.1051/0004-6361/201833051}, \href
  {https://ui.adsabs.harvard.edu/abs/2018A&A...616A...1G} {616, A1}

\bibitem[\protect\citeauthoryear{{Gaia Collaboration} et~al.,}{{Gaia
  Collaboration} et~al.}{2021}]{2021A&A...649A...1G}
{Gaia Collaboration} et~al., 2021, \mn@doi [\aap]
  {10.1051/0004-6361/202039657}, \href
  {https://ui.adsabs.harvard.edu/abs/2021A&A...649A...1G} {649, A1}

\bibitem[\protect\citeauthoryear{{Gaia Collaboration} et~al.,}{{Gaia
  Collaboration} et~al.}{2022}]{2022arXiv220800211G}
{Gaia Collaboration} et~al., 2022, \mn@doi [arXiv e-prints]
  {10.48550/arXiv.2208.00211}, \href
  {https://ui.adsabs.harvard.edu/abs/2022arXiv220800211G} {p. arXiv:2208.00211}

\bibitem[\protect\citeauthoryear{{Gaidos}, {Claytor}, {Dungee}, {Ali}  \&
  {Feiden}}{{Gaidos} et~al.}{2023}]{2023MNRAS.520.5283G}
{Gaidos} E.,  {Claytor} Z.,  {Dungee} R.,  {Ali} A.,   {Feiden} G.~A.,  2023,
  \mn@doi [\mnras] {10.1093/mnras/stad343}, \href
  {https://ui.adsabs.harvard.edu/abs/2023MNRAS.520.5283G} {520, 5283}

\bibitem[\protect\citeauthoryear{{Garc{\'\i}a} et~al.,}{{Garc{\'\i}a}
  et~al.}{2014}]{2014A&A...572A..34G}
{Garc{\'\i}a} R.~A.,  et~al., 2014, \mn@doi [\aap]
  {10.1051/0004-6361/201423888}, \href
  {https://ui.adsabs.harvard.edu/abs/2014A&A...572A..34G} {572, A34}

\bibitem[\protect\citeauthoryear{{Giles}, {Collier Cameron}  \&
  {Haywood}}{{Giles} et~al.}{2017}]{2017MNRAS.472.1618G}
{Giles} H. A.~C.,  {Collier Cameron} A.,   {Haywood} R.~D.,  2017, \mn@doi
  [\mnras] {10.1093/mnras/stx1931}, \href
  {https://ui.adsabs.harvard.edu/abs/2017MNRAS.472.1618G} {472, 1618}

\bibitem[\protect\citeauthoryear{{Godoy-Rivera} \&
  {Chanam{\'e}}}{{Godoy-Rivera} \& {Chanam{\'e}}}{2018}]{2018MNRAS.479.4440G}
{Godoy-Rivera} D.,  {Chanam{\'e}} J.,  2018, \mn@doi [\mnras]
  {10.1093/mnras/sty1736}, \href
  {https://ui.adsabs.harvard.edu/abs/2018MNRAS.479.4440G} {479, 4440}

\bibitem[\protect\citeauthoryear{{Godoy-Rivera}, {Pinsonneault}  \&
  {Rebull}}{{Godoy-Rivera} et~al.}{2021a}]{2021ApJS..257...46G}
{Godoy-Rivera} D.,  {Pinsonneault} M.~H.,   {Rebull} L.~M.,  2021a, \mn@doi
  [\apjs] {10.3847/1538-4365/ac2058}, \href
  {https://ui.adsabs.harvard.edu/abs/2021ApJS..257...46G} {257, 46}

\bibitem[\protect\citeauthoryear{{Godoy-Rivera} et~al.,}{{Godoy-Rivera}
  et~al.}{2021b}]{2021ApJ...915...19G}
{Godoy-Rivera} D.,  et~al., 2021b, \mn@doi [\apj] {10.3847/1538-4357/abf8ba},
  \href {https://ui.adsabs.harvard.edu/abs/2021ApJ...915...19G} {915, 19}

\bibitem[\protect\citeauthoryear{{Gordon}, {Davenport}, {Angus},
  {Foreman-Mackey}, {Agol}, {Covey}, {Ag{\"u}eros}  \& {Kipping}}{{Gordon}
  et~al.}{2021}]{2021ApJ...913...70G}
{Gordon} T.~A.,  {Davenport} J. R.~A.,  {Angus} R.,  {Foreman-Mackey} D.,
  {Agol} E.,  {Covey} K.~R.,  {Ag{\"u}eros} M.~A.,   {Kipping} D.,  2021,
  \mn@doi [\apj] {10.3847/1538-4357/abf63e}, \href
  {https://ui.adsabs.harvard.edu/abs/2021ApJ...913...70G} {913, 70}

\bibitem[\protect\citeauthoryear{{Green}, {Schlafly}, {Zucker}, {Speagle}  \&
  {Finkbeiner}}{{Green} et~al.}{2019}]{2019ApJ...887...93G}
{Green} G.~M.,  {Schlafly} E.,  {Zucker} C.,  {Speagle} J.~S.,   {Finkbeiner}
  D.,  2019, \mn@doi [\apj] {10.3847/1538-4357/ab5362}, \href
  {https://ui.adsabs.harvard.edu/abs/2019ApJ...887...93G} {887, 93}

\bibitem[\protect\citeauthoryear{{Gruner} \& {Barnes}}{{Gruner} \&
  {Barnes}}{2020}]{2020A&A...644A..16G}
{Gruner} D.,  {Barnes} S.~A.,  2020, \mn@doi [\aap]
  {10.1051/0004-6361/202038984}, \href
  {https://ui.adsabs.harvard.edu/abs/2020A&A...644A..16G} {644, A16}

\bibitem[\protect\citeauthoryear{{Hall} et~al.,}{{Hall}
  et~al.}{2021}]{2021NatAs...5..707H}
{Hall} O.~J.,  et~al., 2021, \mn@doi [Nature Astronomy]
  {10.1038/s41550-021-01335-x}, \href
  {https://ui.adsabs.harvard.edu/abs/2021NatAs...5..707H} {5, 707}

\bibitem[\protect\citeauthoryear{{Hartman} \& {L{\'e}pine}}{{Hartman} \&
  {L{\'e}pine}}{2020}]{2020ApJS..247...66H}
{Hartman} Z.~D.,  {L{\'e}pine} S.,  2020, \mn@doi [\apjs]
  {10.3847/1538-4365/ab79a6}, \href
  {https://ui.adsabs.harvard.edu/abs/2020ApJS..247...66H} {247, 66}

\bibitem[\protect\citeauthoryear{{Howell} et~al.,}{{Howell}
  et~al.}{2014}]{2014PASP..126..398H}
{Howell} S.~B.,  et~al., 2014, \mn@doi [\pasp] {10.1086/676406}, \href
  {https://ui.adsabs.harvard.edu/abs/2014PASP..126..398H} {126, 398}

\bibitem[\protect\citeauthoryear{{Hwang}, {Ting}  \& {Zakamska}}{{Hwang}
  et~al.}{2022}]{2022MNRAS.512.3383H}
{Hwang} H.-C.,  {Ting} Y.-S.,   {Zakamska} N.~L.,  2022, \mn@doi [\mnras]
  {10.1093/mnras/stac675}, \href
  {https://ui.adsabs.harvard.edu/abs/2022MNRAS.512.3383H} {512, 3383}

\bibitem[\protect\citeauthoryear{{Ilic}, {Poppenhaeger}  \& {Hosseini}}{{Ilic}
  et~al.}{2022}]{2022MNRAS.513.4380I}
{Ilic} N.,  {Poppenhaeger} K.,   {Hosseini} S.~M.,  2022, \mn@doi [\mnras]
  {10.1093/mnras/stac861}, \href
  {https://ui.adsabs.harvard.edu/abs/2022MNRAS.513.4380I} {513, 4380}

\bibitem[\protect\citeauthoryear{{Janes}}{{Janes}}{2017}]{2017ApJ...835...75J}
{Janes} K.~A.,  2017, \mn@doi [\apj] {10.3847/1538-4357/835/1/75}, \href
  {https://ui.adsabs.harvard.edu/abs/2017ApJ...835...75J} {835, 75}

\bibitem[\protect\citeauthoryear{{Jordi} et~al.,}{{Jordi}
  et~al.}{2010}]{2010A&A...523A..48J}
{Jordi} C.,  et~al., 2010, \mn@doi [\aap] {10.1051/0004-6361/201015441}, \href
  {https://ui.adsabs.harvard.edu/abs/2010A&A...523A..48J} {523, A48}

\bibitem[\protect\citeauthoryear{{Katz} et~al.,}{{Katz}
  et~al.}{2022}]{2022arXiv220605902K}
{Katz} D.,  et~al., 2022, \mn@doi [arXiv e-prints] {10.48550/arXiv.2206.05902},
  \href {https://ui.adsabs.harvard.edu/abs/2022arXiv220605902K} {p.
  arXiv:2206.05902}

\bibitem[\protect\citeauthoryear{{Kawaler}}{{Kawaler}}{1988}]{1988ApJ...333..236K}
{Kawaler} S.~D.,  1988, \mn@doi [\apj] {10.1086/166740}, \href
  {https://ui.adsabs.harvard.edu/abs/1988ApJ...333..236K} {333, 236}

\bibitem[\protect\citeauthoryear{{Kounkel}, {Stassun}, {Bouma}, {Covey},
  {Hillenbrand}  \& {Curtis}}{{Kounkel} et~al.}{2022}]{2022AJ....164..137K}
{Kounkel} M.,  {Stassun} K.~G.,  {Bouma} L.~G.,  {Covey} K.,  {Hillenbrand}
  L.~A.,   {Curtis} J.~L.,  2022, \mn@doi [\aj] {10.3847/1538-3881/ac866d},
  \href {https://ui.adsabs.harvard.edu/abs/2022AJ....164..137K} {164, 137}

\bibitem[\protect\citeauthoryear{{Kouwenhoven}, {Goodwin}, {Parker}, {Davies},
  {Malmberg}  \& {Kroupa}}{{Kouwenhoven} et~al.}{2010}]{2010MNRAS.404.1835K}
{Kouwenhoven} M.~B.~N.,  {Goodwin} S.~P.,  {Parker} R.~J.,  {Davies} M.~B.,
  {Malmberg} D.,   {Kroupa} P.,  2010, \mn@doi [\mnras]
  {10.1111/j.1365-2966.2010.16399.x}, \href
  {https://ui.adsabs.harvard.edu/abs/2010MNRAS.404.1835K} {404, 1835}

\bibitem[\protect\citeauthoryear{{Kraus} \& {Hillenbrand}}{{Kraus} \&
  {Hillenbrand}}{2009}]{2009ApJ...704..531K}
{Kraus} A.~L.,  {Hillenbrand} L.~A.,  2009, \mn@doi [\apj]
  {10.1088/0004-637X/704/1/531}, \href
  {https://ui.adsabs.harvard.edu/abs/2009ApJ...704..531K} {704, 531}

\bibitem[\protect\citeauthoryear{{Lee}, {Lee}, {Dunham}, {Tatematsu}, {Choi},
  {Bergin}  \& {Evans}}{{Lee} et~al.}{2017}]{2017NatAs...1E.172L}
{Lee} J.-E.,  {Lee} S.,  {Dunham} M.~M.,  {Tatematsu} K.,  {Choi} M.,  {Bergin}
  E.~A.,   {Evans} N.~J.,  2017, \mn@doi [Nature Astronomy]
  {10.1038/s41550-017-0172}, \href
  {https://ui.adsabs.harvard.edu/abs/2017NatAs...1E.172L} {1, 0172}

\bibitem[\protect\citeauthoryear{{L{\'e}pine} \& {Bongiorno}}{{L{\'e}pine} \&
  {Bongiorno}}{2007}]{2007AJ....133..889L}
{L{\'e}pine} S.,  {Bongiorno} B.,  2007, \mn@doi [\aj] {10.1086/510333}, \href
  {https://ui.adsabs.harvard.edu/abs/2007AJ....133..889L} {133, 889}

\bibitem[\protect\citeauthoryear{{Lu}, {Angus}, {Curtis}, {David}  \&
  {Kiman}}{{Lu} et~al.}{2021}]{2021AJ....161..189L}
{Lu} Y.~L.,  {Angus} R.,  {Curtis} J.~L.,  {David} T.~J.,   {Kiman} R.,  2021,
  \mn@doi [\aj] {10.3847/1538-3881/abe4d6}, \href
  {https://ui.adsabs.harvard.edu/abs/2021AJ....161..189L} {161, 189}

\bibitem[\protect\citeauthoryear{{Lu}, {Curtis}, {Angus}, {David}  \&
  {Hattori}}{{Lu} et~al.}{2022}]{2022AJ....164..251L}
{Lu} Y.~L.,  {Curtis} J.~L.,  {Angus} R.,  {David} T.~J.,   {Hattori} S.,
  2022, \mn@doi [\aj] {10.3847/1538-3881/ac9bee}, \href
  {https://ui.adsabs.harvard.edu/abs/2022AJ....164..251L} {164, 251}

\bibitem[\protect\citeauthoryear{{Mamajek} \& {Hillenbrand}}{{Mamajek} \&
  {Hillenbrand}}{2008}]{2008ApJ...687.1264M}
{Mamajek} E.~E.,  {Hillenbrand} L.~A.,  2008, \mn@doi [\apj] {10.1086/591785},
  \href {https://ui.adsabs.harvard.edu/abs/2008ApJ...687.1264M} {687, 1264}

\bibitem[\protect\citeauthoryear{{Masuda}, {Petigura}  \& {Hall}}{{Masuda}
  et~al.}{2022}]{2022MNRAS.510.5623M}
{Masuda} K.,  {Petigura} E.~A.,   {Hall} O.~J.,  2022, \mn@doi [\mnras]
  {10.1093/mnras/stab3650}, \href
  {https://ui.adsabs.harvard.edu/abs/2022MNRAS.510.5623M} {510, 5623}

\bibitem[\protect\citeauthoryear{{Mathur} et~al.,}{{Mathur}
  et~al.}{2022}]{2022A&A...657A..31M}
{Mathur} S.,  et~al., 2022, \mn@doi [\aap] {10.1051/0004-6361/202141168}, \href
  {https://ui.adsabs.harvard.edu/abs/2022A&A...657A..31M} {657, A31}

\bibitem[\protect\citeauthoryear{{Maxted}, {Serenelli}  \&
  {Southworth}}{{Maxted} et~al.}{2015}]{2015A&A...577A..90M}
{Maxted} P.~F.~L.,  {Serenelli} A.~M.,   {Southworth} J.,  2015, \mn@doi [\aap]
  {10.1051/0004-6361/201525774}, \href
  {https://ui.adsabs.harvard.edu/abs/2015A&A...577A..90M} {577, A90}

\bibitem[\protect\citeauthoryear{{McQuillan}, {Mazeh}  \&
  {Aigrain}}{{McQuillan} et~al.}{2014}]{2014ApJS..211...24M}
{McQuillan} A.,  {Mazeh} T.,   {Aigrain} S.,  2014, \mn@doi [\apjs]
  {10.1088/0067-0049/211/2/24}, \href
  {https://ui.adsabs.harvard.edu/abs/2014ApJS..211...24M} {211, 24}

\bibitem[\protect\citeauthoryear{{Meibom}, {Mathieu}  \& {Stassun}}{{Meibom}
  et~al.}{2009}]{2009ApJ...695..679M}
{Meibom} S.,  {Mathieu} R.~D.,   {Stassun} K.~G.,  2009, \mn@doi [\apj]
  {10.1088/0004-637X/695/1/679}, \href
  {https://ui.adsabs.harvard.edu/abs/2009ApJ...695..679M} {695, 679}

\bibitem[\protect\citeauthoryear{{Meibom} et~al.,}{{Meibom}
  et~al.}{2011}]{2011ApJ...733L...9M}
{Meibom} S.,  et~al., 2011, \mn@doi [\apjl] {10.1088/2041-8205/733/1/L9}, \href
  {https://ui.adsabs.harvard.edu/abs/2011ApJ...733L...9M} {733, L9}

\bibitem[\protect\citeauthoryear{{Meibom}, {Barnes}, {Platais}, {Gilliland},
  {Latham}  \& {Mathieu}}{{Meibom} et~al.}{2015}]{2015Natur.517..589M}
{Meibom} S.,  {Barnes} S.~A.,  {Platais} I.,  {Gilliland} R.~L.,  {Latham}
  D.~W.,   {Mathieu} R.~D.,  2015, \mn@doi [\nat] {10.1038/nature14118}, \href
  {https://ui.adsabs.harvard.edu/abs/2015Natur.517..589M} {517, 589}

\bibitem[\protect\citeauthoryear{{Messina}, {Nardiello}, {Desidera},
  {Baratella}, {Benatti}, {Biazzo}  \& {D'Orazi}}{{Messina}
  et~al.}{2022}]{2022A&A...657L...3M}
{Messina} S.,  {Nardiello} D.,  {Desidera} S.,  {Baratella} M.,  {Benatti} S.,
  {Biazzo} K.,   {D'Orazi} V.,  2022, \mn@doi [\aap]
  {10.1051/0004-6361/202142276}, \href
  {https://ui.adsabs.harvard.edu/abs/2022A&A...657L...3M} {657, L3}

\bibitem[\protect\citeauthoryear{{Metcalfe} \& {Egeland}}{{Metcalfe} \&
  {Egeland}}{2019}]{2019ApJ...871...39M}
{Metcalfe} T.~S.,  {Egeland} R.,  2019, \mn@doi [\apj]
  {10.3847/1538-4357/aaf575}, \href
  {https://ui.adsabs.harvard.edu/abs/2019ApJ...871...39M} {871, 39}

\bibitem[\protect\citeauthoryear{{Moeckel} \& {Clarke}}{{Moeckel} \&
  {Clarke}}{2011}]{2011MNRAS.415.1179M}
{Moeckel} N.,  {Clarke} C.~J.,  2011, \mn@doi [\mnras]
  {10.1111/j.1365-2966.2011.18731.x}, \href
  {https://ui.adsabs.harvard.edu/abs/2011MNRAS.415.1179M} {415, 1179}

\bibitem[\protect\citeauthoryear{{Ness} et~al.,}{{Ness}
  et~al.}{2018}]{2018ApJ...853..198N}
{Ness} M.,  et~al., 2018, \mn@doi [\apj] {10.3847/1538-4357/aa9d8e}, \href
  {https://ui.adsabs.harvard.edu/abs/2018ApJ...853..198N} {853, 198}

\bibitem[\protect\citeauthoryear{{Newton}, {Irwin}, {Charbonneau},
  {Berta-Thompson}, {Dittmann}  \& {West}}{{Newton}
  et~al.}{2016}]{2016ApJ...821...93N}
{Newton} E.~R.,  {Irwin} J.,  {Charbonneau} D.,  {Berta-Thompson} Z.~K.,
  {Dittmann} J.~A.,   {West} A.~A.,  2016, \mn@doi [\apj]
  {10.3847/0004-637X/821/2/93}, \href
  {https://ui.adsabs.harvard.edu/abs/2016ApJ...821...93N} {821, 93}

\bibitem[\protect\citeauthoryear{{Offner}, {Kratter}, {Matzner}, {Krumholz}  \&
  {Klein}}{{Offner} et~al.}{2010}]{2010ApJ...725.1485O}
{Offner} S. S.~R.,  {Kratter} K.~M.,  {Matzner} C.~D.,  {Krumholz} M.~R.,
  {Klein} R.~I.,  2010, \mn@doi [\apj] {10.1088/0004-637X/725/2/1485}, \href
  {https://ui.adsabs.harvard.edu/abs/2010ApJ...725.1485O} {725, 1485}

\bibitem[\protect\citeauthoryear{{Otani}, {von Hippel}, {Buzasi}, {Oswalt},
  {Stone-Martinez}  \& {Majewski}}{{Otani} et~al.}{2022}]{2022ApJ...930...36O}
{Otani} T.,  {von Hippel} T.,  {Buzasi} D.,  {Oswalt} T.~D.,  {Stone-Martinez}
  A.,   {Majewski} P.,  2022, \mn@doi [\apj] {10.3847/1538-4357/ac6035}, \href
  {https://ui.adsabs.harvard.edu/abs/2022ApJ...930...36O} {930, 36}

\bibitem[\protect\citeauthoryear{{Parker}}{{Parker}}{1958}]{1958ApJ...128..664P}
{Parker} E.~N.,  1958, \mn@doi [\apj] {10.1086/146579}, \href
  {https://ui.adsabs.harvard.edu/abs/1958ApJ...128..664P} {128, 664}

\bibitem[\protect\citeauthoryear{{Pass}, {Charbonneau}, {Irwin}  \&
  {Winters}}{{Pass} et~al.}{2022}]{2022ApJ...936..109P}
{Pass} E.~K.,  {Charbonneau} D.,  {Irwin} J.~M.,   {Winters} J.~G.,  2022,
  \mn@doi [\apj] {10.3847/1538-4357/ac7da8}, \href
  {https://ui.adsabs.harvard.edu/abs/2022ApJ...936..109P} {936, 109}

\bibitem[\protect\citeauthoryear{{Pe{\~n}arrubia}}{{Pe{\~n}arrubia}}{2021}]{2021MNRAS.501.3670P}
{Pe{\~n}arrubia} J.,  2021, \mn@doi [\mnras] {10.1093/mnras/staa3700}, \href
  {https://ui.adsabs.harvard.edu/abs/2021MNRAS.501.3670P} {501, 3670}

\bibitem[\protect\citeauthoryear{{Pinsonneault} et~al.,}{{Pinsonneault}
  et~al.}{2018}]{2018ApJS..239...32P}
{Pinsonneault} M.~H.,  et~al., 2018, \mn@doi [\apjs]
  {10.3847/1538-4365/aaebfd}, \href
  {https://ui.adsabs.harvard.edu/abs/2018ApJS..239...32P} {239, 32}

\bibitem[\protect\citeauthoryear{{Ram{\'\i}rez}, {Khanal}, {Lichon},
  {Chanam{\'e}}, {Endl}, {Mel{\'e}ndez}  \& {Lambert}}{{Ram{\'\i}rez}
  et~al.}{2019}]{2019MNRAS.490.2448R}
{Ram{\'\i}rez} I.,  {Khanal} S.,  {Lichon} S.~J.,  {Chanam{\'e}} J.,  {Endl}
  M.,  {Mel{\'e}ndez} J.,   {Lambert} D.~L.,  2019, \mn@doi [\mnras]
  {10.1093/mnras/stz2709}, \href
  {https://ui.adsabs.harvard.edu/abs/2019MNRAS.490.2448R} {490, 2448}

\bibitem[\protect\citeauthoryear{{Rebull}, {Stauffer}, {Hillenbrand}, {Cody},
  {Bouvier}, {Soderblom}, {Pinsonneault}  \& {Hebb}}{{Rebull}
  et~al.}{2017}]{2017ApJ...839...92R}
{Rebull} L.~M.,  {Stauffer} J.~R.,  {Hillenbrand} L.~A.,  {Cody} A.~M.,
  {Bouvier} J.,  {Soderblom} D.~R.,  {Pinsonneault} M.,   {Hebb} L.,  2017,
  \mn@doi [\apj] {10.3847/1538-4357/aa6aa4}, \href
  {https://ui.adsabs.harvard.edu/abs/2017ApJ...839...92R} {839, 92}

\bibitem[\protect\citeauthoryear{{Reinhold} \& {Hekker}}{{Reinhold} \&
  {Hekker}}{2020}]{2020A&A...635A..43R}
{Reinhold} T.,  {Hekker} S.,  2020, \mn@doi [\aap]
  {10.1051/0004-6361/201936887}, \href
  {https://ui.adsabs.harvard.edu/abs/2020A&A...635A..43R} {635, A43}

\bibitem[\protect\citeauthoryear{{Sahlholdt}, {Feltzing}  \&
  {Feuillet}}{{Sahlholdt} et~al.}{2022}]{2022MNRAS.510.4669S}
{Sahlholdt} C.~L.,  {Feltzing} S.,   {Feuillet} D.~K.,  2022, \mn@doi [\mnras]
  {10.1093/mnras/stab3681}, \href
  {https://ui.adsabs.harvard.edu/abs/2022MNRAS.510.4669S} {510, 4669}

\bibitem[\protect\citeauthoryear{{Santos}, {Garc{\'\i}a}, {Mathur}, {Bugnet},
  {van Saders}, {Metcalfe}, {Simonian}  \& {Pinsonneault}}{{Santos}
  et~al.}{2019}]{2019ApJS..244...21S}
{Santos} A.~R.~G.,  {Garc{\'\i}a} R.~A.,  {Mathur} S.,  {Bugnet} L.,  {van
  Saders} J.~L.,  {Metcalfe} T.~S.,  {Simonian} G.~V.~A.,   {Pinsonneault}
  M.~H.,  2019, \mn@doi [\apjs] {10.3847/1538-4365/ab3b56}, \href
  {https://ui.adsabs.harvard.edu/abs/2019ApJS..244...21S} {244, 21}

\bibitem[\protect\citeauthoryear{{Santos}, {Breton}, {Mathur}  \&
  {Garc{\'\i}a}}{{Santos} et~al.}{2021a}]{2021ApJS..255...17S}
{Santos} A.~R.~G.,  {Breton} S.~N.,  {Mathur} S.,   {Garc{\'\i}a} R.~A.,
  2021a, \mn@doi [\apjs] {10.3847/1538-4365/ac033f}, \href
  {https://ui.adsabs.harvard.edu/abs/2021ApJS..255...17S} {255, 17}

\bibitem[\protect\citeauthoryear{{Santos}, {Mathur}, {Garc{\'\i}a}, {Cunha}  \&
  {Avelino}}{{Santos} et~al.}{2021b}]{2021MNRAS.508..267S}
{Santos} A.~R.~G.,  {Mathur} S.,  {Garc{\'\i}a} R.~A.,  {Cunha} M.~S.,
  {Avelino} P.~P.,  2021b, \mn@doi [\mnras] {10.1093/mnras/stab2402}, \href
  {https://ui.adsabs.harvard.edu/abs/2021MNRAS.508..267S} {508, 267}

\bibitem[\protect\citeauthoryear{{Schatzman}}{{Schatzman}}{1962}]{1962AnAp...25...18S}
{Schatzman} E.,  1962, Annales d'Astrophysique, \href
  {https://ui.adsabs.harvard.edu/abs/1962AnAp...25...18S} {25, 18}

\bibitem[\protect\citeauthoryear{{Simonian}, {Pinsonneault}  \&
  {Terndrup}}{{Simonian} et~al.}{2019}]{2019ApJ...871..174S}
{Simonian} G. V.~A.,  {Pinsonneault} M.~H.,   {Terndrup} D.~M.,  2019, \mn@doi
  [\apj] {10.3847/1538-4357/aaf97c}, \href
  {https://ui.adsabs.harvard.edu/abs/2019ApJ...871..174S} {871, 174}

\bibitem[\protect\citeauthoryear{{Simonian}, {Pinsonneault}, {Terndrup}  \&
  {van Saders}}{{Simonian} et~al.}{2020}]{2020ApJ...898...76S}
{Simonian} G. V.~A.,  {Pinsonneault} M.~H.,  {Terndrup} D.~M.,   {van Saders}
  J.~L.,  2020, \mn@doi [\apj] {10.3847/1538-4357/ab9a43}, \href
  {https://ui.adsabs.harvard.edu/abs/2020ApJ...898...76S} {898, 76}

\bibitem[\protect\citeauthoryear{{Skumanich}}{{Skumanich}}{1972}]{1972ApJ...171..565S}
{Skumanich} A.,  1972, \mn@doi [\apj] {10.1086/151310}, \href
  {https://ui.adsabs.harvard.edu/abs/1972ApJ...171..565S} {171, 565}

\bibitem[\protect\citeauthoryear{{Soderblom}}{{Soderblom}}{2010}]{2010ARA&A..48..581S}
{Soderblom} D.~R.,  2010, \mn@doi [\araa]
  {10.1146/annurev-astro-081309-130806}, \href
  {https://ui.adsabs.harvard.edu/abs/2010ARA&A..48..581S} {48, 581}

\bibitem[\protect\citeauthoryear{{Sollima}, {Carballo-Bello}, {Beccari},
  {Ferraro}, {Pecci}  \& {Lanzoni}}{{Sollima}
  et~al.}{2010}]{2010MNRAS.401..577S}
{Sollima} A.,  {Carballo-Bello} J.~A.,  {Beccari} G.,  {Ferraro} F.~R.,
  {Pecci} F.~F.,   {Lanzoni} B.,  2010, \mn@doi [\mnras]
  {10.1111/j.1365-2966.2009.15676.x}, \href
  {https://ui.adsabs.harvard.edu/abs/2010MNRAS.401..577S} {401, 577}

\bibitem[\protect\citeauthoryear{{Spada} \& {Lanzafame}}{{Spada} \&
  {Lanzafame}}{2020}]{2020A&A...636A..76S}
{Spada} F.,  {Lanzafame} A.~C.,  2020, \mn@doi [\aap]
  {10.1051/0004-6361/201936384}, \href
  {https://ui.adsabs.harvard.edu/abs/2020A&A...636A..76S} {636, A76}

\bibitem[\protect\citeauthoryear{{Turner} et~al.,}{{Turner}
  et~al.}{2022}]{2022MNRAS.516.4612T}
{Turner} J.~A.,  et~al., 2022, \mn@doi [\mnras] {10.1093/mnras/stac2559}, \href
  {https://ui.adsabs.harvard.edu/abs/2022MNRAS.516.4612T} {516, 4612}

\bibitem[\protect\citeauthoryear{{Xiang} \& {Rix}}{{Xiang} \&
  {Rix}}{2022}]{2022Natur.603..599X}
{Xiang} M.,  {Rix} H.-W.,  2022, \mn@doi [\nat] {10.1038/s41586-022-04496-5},
  \href {https://ui.adsabs.harvard.edu/abs/2022Natur.603..599X} {603, 599}

\bibitem[\protect\citeauthoryear{{Xiang} et~al.,}{{Xiang}
  et~al.}{2017}]{2017ApJS..232....2X}
{Xiang} M.,  et~al., 2017, \mn@doi [\apjs] {10.3847/1538-4365/aa80e4}, \href
  {https://ui.adsabs.harvard.edu/abs/2017ApJS..232....2X} {232, 2}

\bibitem[\protect\citeauthoryear{{do Nascimento} J.~D. et~al.,}{{do Nascimento}
  et~al.}{2014}]{2014ApJ...790L..23D}
{do Nascimento} J.~D. J.,  et~al., 2014, \mn@doi [\apjl]
  {10.1088/2041-8205/790/2/L23}, \href
  {https://ui.adsabs.harvard.edu/abs/2014ApJ...790L..23D} {790, L23}

\bibitem[\protect\citeauthoryear{{van Leeuwen} et~al.,}{{van Leeuwen}
  et~al.}{2018}]{2018gdr2.reptE....V}
{van Leeuwen} F.,  et~al., 2018, {Gaia DR2 documentation}, Gaia DR2
  documentation, European Space Agency; Gaia Data Processing and Analysis
  Consortium. Online at <A
  href=``https://gea.esac.esa.int/archive/documentation/GDR2/''>https://gea.esac.esa.int/archive/documentation/GDR2/</A>

\bibitem[\protect\citeauthoryear{{van Saders}, {Ceillier}, {Metcalfe}, {Silva
  Aguirre}, {Pinsonneault}, {Garc{\'\i}a}, {Mathur}  \& {Davies}}{{van Saders}
  et~al.}{2016}]{2016Natur.529..181V}
{van Saders} J.~L.,  {Ceillier} T.,  {Metcalfe} T.~S.,  {Silva Aguirre} V.,
  {Pinsonneault} M.~H.,  {Garc{\'\i}a} R.~A.,  {Mathur} S.,   {Davies} G.~R.,
  2016, \mn@doi [\nat] {10.1038/nature16168}, \href
  {https://ui.adsabs.harvard.edu/abs/2016Natur.529..181V} {529, 181}

\bibitem[\protect\citeauthoryear{{van Saders}, {Pinsonneault}  \&
  {Barbieri}}{{van Saders} et~al.}{2019}]{2019ApJ...872..128V}
{van Saders} J.~L.,  {Pinsonneault} M.~H.,   {Barbieri} M.,  2019, \mn@doi
  [\apj] {10.3847/1538-4357/aafafe}, \href
  {https://ui.adsabs.harvard.edu/abs/2019ApJ...872..128V} {872, 128}

\makeatother
\end{thebibliography}




\appendix

\section{Description of data table} \label{sec:data_table}

Table \ref{tab:data_table_description} describes the data tables available in the linked repository. In the full data tables, we list the values of $\Delta P_{rot,gyro}$, $\sigma_{\Delta P_{rot,gyro}}$ and $x$ for the pairs, as well as rotation period and other astrometric and photometric data for the binary components. As for the Clusters and Random Pairs [1] samples we did a process of x20 over-sampling to increase their sizes, the table also lists the number of iteration of said over-sampling (thus taking values between 1 and 20). Note that three data tables are provided, one containing the pairs used for testing the \citet{2019AJ....158..173A} relation, one for the \citet{2015MNRAS.450.1787A} relation, and one for \citet{2020A&A...636A..76S}.

\begin{table*}
    \centering
    \begin{tabular}{lll}
    \hline
    Column & Units & Description \\ 
    \hline
    flag\_sample$^a$   & - & Sample that the pair corresponds to\\
    Delta\_P\_rot\_gyro   & days & $\Delta P_{rot,gyro}$ given by Equation (\ref{eq:gyro}), for the pair\\
    sigma\_Delta\_P\_rot\_gyro  & days & Uncertainty of $\Delta P_{rot,gyro}$, as described in Section \ref{sec:rot_test}, for the pair\\
    x\_parameter  & - & x-parameter given by Equation (\ref{eq:x}), for the pair \\
    Delta\_theta  & arcsec & Angular separation of the pair \\
    Delta\_mu  & mas yr$^{-1}$ & Proper motion difference of the pair\\
    Source\_star\_1$^b$  & - & Source of the star in the pair, for the primary\\
    Kepler\_ID\_1  & - & Source ID of the \textit{Kepler} mission (either KIC or EPIC) for the primary\\
    Source\_P\_rot\_1$^b$  & - & Source of the rotation period of the primary\\
    P\_rot\_1  & days & Rotation period of the primary \\
    e\_P\_rot\_1  & days & Uncertainty of the rotation period of the primary \\
    Age\_1  & yr & Age calculated with the corresponding gyrochronology relation,$^c$ of the primary\\
    Gaia\_DR2\_ID\_1  & - & Gaia DR2 Source ID of the primary\\
    ra\_J20155\_1  & deg & R.A. in ICRS Epoch 2015.5, for the primary\\
    dec\_J20155\_1  & deg & Declination in ICRS Epoch 2015.5, for the primary\\
    pmra\_1  & mas yr$^{-1}$ & Proper motion in R.A., for the primary\\
    pmdec\_1  & mas yr$^{-1}$ & Proper motion in declination, for the primary\\
    parallax\_1  & mas & Parallax for the primary\\
    distance\_1  & pc & Distance given by \citet{2018AJ....156...58B}, for the primary\\
    Bayestar19\_1  & mag & Bayestar19 extinction parameter, for the primary\\
    A\_V\_1  & mag & V band extinction parameter, for the primary\\
    A\_G\_1  & mag & Gaia $G$ band extinction parameter, for the primary\\
    E\_BPRP\_1  & mag & Gaia $G_{BP} - G_{RP}$ color reddening parameter, for the primary\\
    G\_apparent\_1  & mag & Apparent Gaia $G$ magnitude, for the primary\\
    G\_0\_absolute\_1  & mag & Absolute, corrected by extinction, Gaia $M_{G0}$ magnitude, for the primary\\
    BPRP\_color\_1  & mag & Unreddened Gaia $G_{BP} - G_{RP}$ color, for the primary\\
    BPRP\_0\_color\_1  & mag & Reddened Gaia $G_{BP} - G_{RP}$ color, for the primary\\
    Source\_star\_2$^b$  & - & Source of the star in the pair, for the secondary\\
    Kepler\_ID\_2  & - & Source ID of the \textit{Kepler} mission (either KIC or EPIC) for the secondary\\
    Source\_P\_rot\_2$^b$  & - & Source of the rotation period of the secondary\\
    P\_rot\_2  & days & Rotation period of the secondary \\
    e\_P\_rot\_2  & days & Uncertainty of the rotation period of the secondary \\
    Age\_2  & yr & Age calculated with the corresponding gyrochronology relation,$^c$ of the secondary\\
    Gaia\_DR2\_ID\_2  & - & Gaia DR2 Source ID of the secondary\\
    ra\_J20155\_2  & deg & R.A. in ICRS Epoch 2015.5, for the secondary\\
    dec\_J20155\_2  & deg & Declination in ICRS Epoch 2015.5, for the secondary\\
    pmra\_2  & mas yr$^{-1}$ & Proper motion in R.A., for the secondary\\
    pmdec\_2  & mas yr$^{-1}$ & Proper motion in declination, for the secondary\\
    parallax\_2  & mas & Parallax for the secondary\\
    distance\_2  & pc & Distance given by \citet{2018AJ....156...58B}, for the secondary\\
    Bayestar19\_2  & mag & Bayestar19 extinction parameter, for the secondary\\
    A\_V\_2  & mag & V band extinction parameter, for the secondary\\
    A\_G\_2  & mag & Gaia $G$ band extinction parameter, for the secondary\\
    E\_BPRP\_2  & mag & Gaia $G_{BP} - G_{RP}$ color reddening parameter, for the secondary\\
    G\_apparent\_2  & mag & Apparent Gaia $G$ magnitude, for the secondary\\
    G\_0\_absolute\_2  & mag & Absolute, corrected by extinction, Gaia $M_{G0}$ magnitude, for the secondary\\
    BPRP\_color\_2  & mag & Unreddened Gaia $G_{BP} - G_{RP}$ color, for the secondary\\
    BPRP\_0\_color\_2  & mag & Reddened Gaia $G_{BP} - G_{RP}$ color, for the secondary\\
    oversampling\_iteration$^d$  & - & Number of iteration that the pair corresponds to\\
    \hline
    \multicolumn{3}{l}{$^a$ Flags are: ``wide\_binaries'' for Wide Binaries, ``clusters'' for Clusters, ``random\_pairs\_1'' for Random Pairs [1], ``random\_pairs\_2'' for Random Pairs [2].} \\
    \multicolumn{3}{l}{\multirow[t]{3}{=}{$^b$ The sources are as follows: (1) \citet{2018MNRAS.479.4440G}; (2) \citet{2017ApJ...835...75J}; (3) \citet{2016MNRAS.455.4212D}; (4) \citet{2018MNRAS.480.4884E}; (5) \citet{2019ApJS..244...21S}; (6) \citet{2014ApJS..211...24M}; (7) Praesepe from \citet{2021ApJS..257...46G}; (8) NGC6811 from \citet{2021ApJS..257...46G}; (9) Ruprecht 147 from \citet{2020A&A...644A..16G}; (10) Ruprecht 147 from \citet{2020ApJ...904..140C}; (11) NGC6819 from \citet{2015Natur.517..589M}.}} \\
    \multicolumn{3}{l}{}\\
    \multicolumn{3}{l}{}\\
    \multicolumn{3}{l}{$^{c}$ Corresponding to the \citet{2019AJ....158..173A}, \citet{2015MNRAS.450.1787A}, or \citet{2020A&A...636A..76S} relations.}\\
    \multicolumn{3}{l}{$^{d}$ Takes values from 1 to 20. For the samples compiled without oversampling (namely, Wide Binaries and Random Pairs [2]), the value is left as blank.}\\
    \hline
    \end{tabular}
    \caption{Description of the columns of the data tables used for this work. The full data tables (one for each gyrochronology relation) are available in the linked repository.}
    \label{tab:data_table_description}
\end{table*}




\bsp	
\label{lastpage}
\end{document}